\documentclass[aps,preprintnumbers,onecolumn,amsmath,amssymb,floatfix,pra,notitlepage,superscriptaddress]{revtex4-1} 

\usepackage{comment}
\usepackage{enumerate}
\usepackage{amssymb}
\usepackage{amsmath}
\usepackage{braket}
\usepackage{graphicx}
\usepackage[usenames,dvipsnames]{color}
\usepackage{tensor}
\usepackage{empheq}
\usepackage{capt-of}
\usepackage[normalem]{ulem} 
\usepackage{soul} %
\usepackage{MnSymbol}

\usepackage[colorlinks,bookmarks=false,citecolor=NavyBlue,linkcolor=OliveGreen,urlcolor=blue]{hyperref}
\newcommand{\be}{\begin{equation}}
\newcommand{\ee}{\end{equation}}
\newcommand{\ba}{\begin{aligned}}
\newcommand{\ea}{\end{aligned}}
\newcommand{\bw}{\begin{widetext}}
\newcommand{\ew}{\end{widetext}}

\newcommand{\bea}{\begin{eqnarray}}
\newcommand{\eea}{\end{eqnarray}}

\def\doi{http://dx.doi.org/}

\begin{document}
\title{Entanglement and classical fluctuations  at finite-temperature critical points}
\author{Sascha Wald}
\affiliation{SISSA and INFN, via Bonomea 265, 34136 Trieste, Italy}
\author{Ra\'ul Arias} 
\affiliation{SISSA and INFN, via Bonomea 265, 34136 Trieste, Italy}
\affiliation{Instituto de F\'isica La Plata - CONICET and Departamento de F\'isica, Universidad Nacional de La Plata C.C. 67, 1900, La Plata, Argentina}
\author{Vincenzo Alba}
\affiliation{Institute  for  Theoretical  Physics, Universiteit van Amsterdam,
Science Park 904, Postbus 94485, 1098 XH Amsterdam,  The  Netherlands}

\begin{abstract}

We investigate several entanglement-related quantities at 
finite-temperature criticality in the three-dimensional quantum spherical model, 
both as a function of temperature $T$ and of the quantum parameter $g$, which 
measures the strength of quantum fluctuations. 
While the von Neumann and the R\'enyi entropies exhibit the volume-law 
for any $g$ and $T$, the mutual information 
obeys an area law. The prefactors of the volume-law and of the area-law are  
regular across the transition, reflecting that universal singular terms 
vanish at the transition. This implies that the mutual 
information is dominated by nonuniversal contributions. This 
hampers its use as a  witness of criticality, at least in the 
spherical model. 
We also study the logarithmic negativity. 
For any value of $g,T$, the negativity exhibits 
an area-law. The negativity vanishes deep in the paramagnetic 
phase, it is larger at 
small temperature,  and it decreases upon increasing the temperature. 
For any $g$, it  
exhibits the so-called sudden death, i.e., it is exactly zero 
for large enough $T$. 
The vanishing of the negativity defines a ``death line'', which 
we characterise analytically at small $g$. 
Importantly, for any finite $T$ the area-law prefactor 
is regular across the transition, whereas it develops a cusp-like 
singularity in the limit $T\to 0$. 
Finally, we consider the single-particle entanglement and 
negativity spectra. The vast majority of 
the levels are regular across the transition. 
Only the larger ones exhibit singularities. These are related 
to the presence of a zero mode, which reflects the  symmetry breaking. 
This implies the presence of sub-leading singular terms in the entanglement 
entropies. Interestingly, since the larger levels do not contribute to the 
negativity, sub-leading singular corrections are expected to be suppressed 
for the negativity.

\end{abstract}

\maketitle

\section{Introduction}

\begin{figure}[t]
\includegraphics[width=0.6\textwidth]{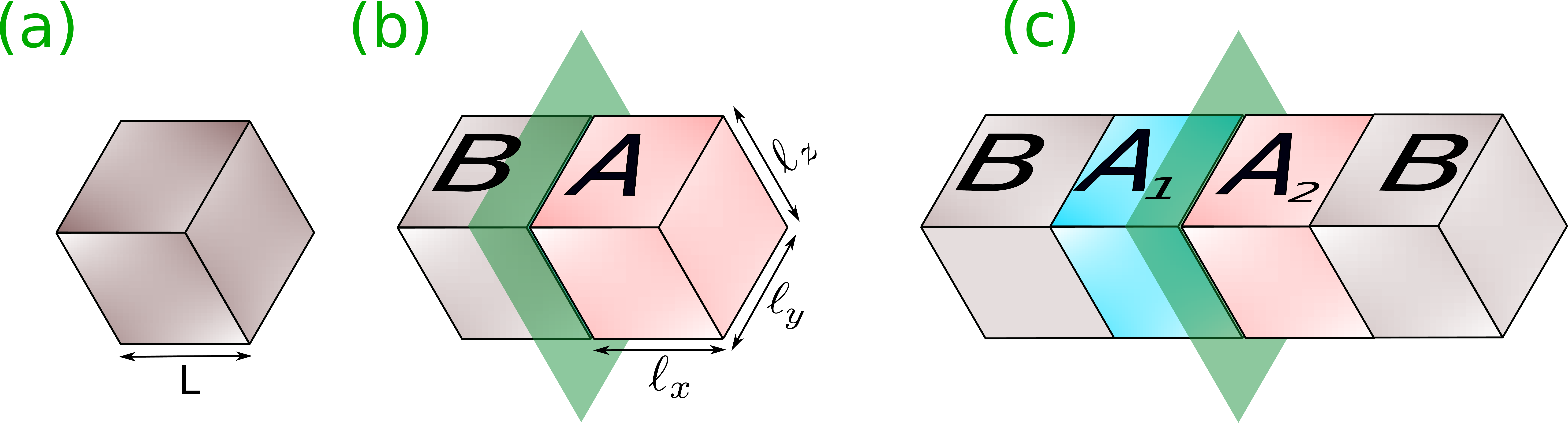}
\caption{ Geometry used in this work. The quantum spherical model 
 lives on a $3D$ cubic lattice of linear size $L$ (see (a)). Periodic 
 boundary conditions are used in all directions. In (b) the total system is 
 divided into two parts as $A\cup B$, and $B$ is traced over. Here 
 $\ell_x\times\ell_y\times\ell_z$ is the volume of $A$. In (c) subsystem 
 $A$ is further divided as $A=A_1\cup A_2$. Here we consider the 
 entanglement between the two adjacent cubic blocks $A_1$ and $A_2$. 
 We consider both the case of finite $B$, as well as the limit of $B$ 
 infinite.  
}
\label{fig:cartoon}
\end{figure}

Understanding the interplay between entanglement and classical fluctuations 
in quantum many-body systems is a challenging task. 
Several entanglement-related quantities have been investigated~\cite{area, amico-2008,
calabrese-2009,laflorencie-2016}. Arguably, 
the most popular ones are the R\'enyi entropies and the 
von Neumann entropy. Given a partition of the system in subsystems 
$A$ and its complement $B\equiv \bar A$ (see Fig.~\ref{fig:cartoon} (a) for 
a three-dimensional setup), the R\'enyi entropies $S_n$ are defined as 
\begin{equation}
	S_n\equiv \frac{1}{1-n}\ln\textrm{Tr}\rho_A^n,
\end{equation}
with $\rho_A$ the reduced density matrix of part $A$. In the limit $n\to1$ 
one obtains the so-called von Neumann entropy $S=-\textrm{Tr}\rho_A\ln\rho_A$. 
If the total system $A\cup \bar A$ is in a pure state, $S_n$ are good measures of 
the entanglement between $A$ and the rest. 
At zero temperature the famous area-law behaviour $S_n\propto L^{d-1}$ holds 
true, with $L$ being the linear size of the system, and $d$ the dimensionality. 
Logarithmic corrections are present for gapless fermionic 
systems~\cite{wolf-2006,gioev-2006,cmv-11d}. 
This is strikingly different at finite temperature, where 
$S_n$ cannot distinguish genuine quantum correlations from thermal ones. 
At finite temperature the volume-law behaviour $S_n\propto L^d$ holds, 
reflecting, for instance, the transformation between entanglement and standard 
thermodynamic entropy. In this situation, the mutual information $I_{n, A:\bar A}$ 
(see section~\ref{sec:obs} 
for its definition) can be used to measure the total correlation, 
i.e., both classical and quantum, between $A$ and $\bar A$.  The 
mutual information exhibits the area-law behaviour $I_n\propto L^{d-1}$. 
Still, $I_n$ is not a proper measure of the mutual entanglement 
between $A$ and $\bar A$. This reflects $A\cup \bar A$ being in a mixed 
state. The so-called logarithmic 
negativity~\cite{lee-2000,vidal-2002,plenio-2005} is the most popular entanglement 
measure for mixed states. For systems described by Conformal Field Theory, 
its behaviour has been fully characterized, both at zero temperature~\cite{calabrese-2012}, 
and at finite temperature~\cite{calabrese-2015}. 
Similar to the mutual information, for any temperature the 
negativity exhibits the area-law scaling, which has been checked 
in several systems~\cite{dct-16,eisler-2016,shapourian-2018,tarun-1,tarun-2}.

An important question, which is key in this paper,  is how quantum 
and classical fluctuations are intertwined at a finite temperature 
critical point. An important remark is that  typical {\it local} 
one dimensional models do not exhibit finite-temperature or mid-spectrum 
criticality, with the notable exception of models that exhibit 
the many-body localisation transition~\cite{nandi}. 
It has been argued that entanglement-related quantities can be used to detect this 
transition~\cite{khemani-17,kjall-2014,chandran-2015,laflorencie-2016}, 
although no conclusion has been reached, due the severe finite-size 
effects. On the other hand, although standard finite-temperature phase transitions 
are possible in higher dimensions, the interplay between 
classical and quantum fluctuations at finite-temperature criticality has  
been scarcely studied, despite the growing interest~\cite{hauke-2016,gabrielli-2018,frerot-2019}. 
Much efforts focused on the behaviour of the mutual information. 
For instance, it has been suggested~\cite{singh-2011,melko-2010,kallin-2013,
inglis-2013,srilu-2018,singh-2012} that the mutual information exhibits 
a crossing for different sizes at a finite-temperature critical 
point, similar to more traditional tools in 
critical phenomena, such as the Binder cumulant~\cite{vicari}. 

Importantly, there is growing interest to understand the 
behaviour of the logarithmic negativity at finite-temperature 
criticality. 
An important question is whether the logarithmic negativity 
exhibits signatures of criticality. 
Since finite-temperature phase transitions are driven by classical 
fluctuations, and the negativity is only sensitive to true quantum correlations, 
one should expect the negativity to be smooth across a finite-temperature 
critical point. Surprisingly, it has been observed that the negativity 
can exhibit ``weak'' singularities, such as cusp-like features, at the 
critical point~\cite{tarun-1,tarun-2,lu-2018}. 
On the other hand, similar singularities have been also observed in 
the area-law prefactor of the R\'enyi entropies at quantum phase 
transitions~\cite{helmes-2014,frerot-2016}.

In this paper we investigate the behaviour of several entanglement-related 
observables at finite-temperature criticality. Specifically, we consider 
the R\'enyi entropies (and the von Neumann entropy), the mutual 
information, and the logarithmic negativity. 
We also discuss the so-called single-particle entanglement spectrum and the 
negativity spectrum. 
To be specific, here we focus on the paradigmatic quantum spherical model 
(QSM) in three dimensions. 
Spherical models have a venerable history, as they served as test ground 
for the theory of critical phenomena and the theory of finite-size 
scaling~\cite{joyce-1972,brankov}. The phase diagram of the model (see Fig.~\ref{fig:phadia}) 
is spanned by the temperature $T$ and the so-called quantum coupling $g$. The latter measures 
the strength of the quantum fluctuations. In three dimensions, the 
model exhibits a line of finite-temperature second-order 
phase transitions between a paramagnetic phase and a ferromagnetically 
ordered phase. The universality class of the transition has been characterised 
analytically, since the model is exactly solvable, and it is that of the 
$N$-vector model~\cite{vicari} in the limit $N\to\infty$ limit~\cite{zinn-1998}. 
Surprisingly, the entanglement properties of spherical models have 
not been explored much. 

First, we verify that the entanglement entropy and 
the R\'enyi entropies exhibit the expected volume-law in the whole 
phase diagram. In the thermodynamic limit, the prefactor of the 
volume-law and its first derivative with respect to the model 
parameters and the temperature are continuous  
across the transition. We refer to this behaviour as {\it regular}. 
Conversely, we refer to non-analytic behaviour as {\it singular}. 
This means that, although the entanglement entropies 
contain singular contributions, these vanish at the transition 
as $|g-g_c|^{\kappa}$ (or as $|T-T_c|^\kappa$) with 
$\kappa>1$, and $T_c$ and $g_c$ marking the critical point. 
The mutual information between two regions exhibits the expected area-law 
behaviour for any value of $g$ and $T$. 
Moreover, although the prefactor of the area law 
contains both regular and singular contributions at criticality, again, 
 we observe regular behaviour across the transition, suggesting that the singular 
terms vanish. 
Interestingly, for finite-size systems we observe that the 
mutual information does not exhibit  a universal crossing at 
the transition.  This reflects that 
the singular terms, which contain universal information, are vanishing at the critical 
point, and the mutual information  is dominated by 
non-universal contributions. 
We find a similar behaviour for the logarithmic negativity. The negativity exhibits 
the expected area-law 
in the whole phase diagram. The area-law prefactor has a maximum at the phase transition. 
In the paramagnetic phase the negativity vanishes as  $1/g$ in the limit $g\to\infty$. 
In the ordered phase it depends mildly on the system 
parameters for low enough temperature and large enough values of $g$. 
For any fixed value of $g$, the negativity  exhibits a ``sudden death'' upon 
increasing the temperature. Specifically, there is a negativity ``death 
line'', above which the negativity is exactly zero. Interestingly,  
in the limit $g\to0$, i.e., when the model becomes classical, the death 
line exhibits the behaviour $\propto g^{1/2}$. 
Importantly, at finite temperature the prefactor of the area-law 
is regular across the para-ferro transition, whereas it 
develops a cusp-like feature upon lowering the temperature. 

Finally, we discuss the single-particle entanglement 
spectrum~\cite{laflorencie-2016} and the negativity spectrum~\cite{ruggiero-2016a}. 
For two nearest-neighbour sites 
in an infinite system, the spectrum contains only two levels. One of the levels 
is regular across the transition, whereas the other exhibits a cusp singularity. 
This is different for the negativity between two 
extended subsystems. Specifically, we observe that the majority of the 
single-particle entanglement spectrum levels show regular behaviour 
at the transition. Only the larger levels exhibit a cusp-like behaviour. 
Interestingly, the corresponding eigenvectors show a flat-in-space 
structure, suggesting that the singular levels originate from 
the zero mode, which is associated with the symmetry breaking. As it is 
well known, this zero mode leads to sub-leading logarithmic terms~\cite{met-grov,alba-2012} 
in the entropies. This also means that, although the prefactor of the 
volume law behaviour of the entropies is regular across the transition, 
the sub-leading corrections exhibit signatures of the criticality. 
We observe a similar structure  for the negativity spectrum. 
However, since the larger negativity spectrum levels do not contribute to the 
logarithmic negativity, this suggests that 
the sub-leading singular terms are suppressed. 

The manuscript is organised as follows. In section~\ref{sec:obs} we introduce the 
entanglement-related quantities that we consider in the paper. In section~\ref{sec:model} 
we review the quantum spherical model. Specifically, in section~\ref{sec-qsm-corr} 
we provide the results for the two-point correlation functions, discussing 
their behaviour in the paramagnetic phase, and in the ordered phase (in section~\ref{sec:large-g} 
and section~\ref{sec:small-g}, respectively). In section~\ref{sec:critical} 
we address the critical behaviour of the model. Criticality 
is encoded in the so-called spherical parameter that we discuss in 
section~\ref{sec:mu}. In section~\ref{sec:mu-crit} we focus on the 
correlators near the para-ferro transition. In order to compare 
entanglement-related quantities with standard thermodynamic ones, in 
section~\ref{sec:free-e} and section~\ref{sec:corr} we discuss 
the singular behaviour of the free energy, and of the universal ratio 
$R_\xi$. In section~\ref{sec-ent-qsm} we focus on the interplay between 
entanglement and finite-temperature criticality. In section~\ref{sec:vn} and 
section~\ref{sec:mi} we study the entanglement entropy, and the 
mutual information. In section~\ref{sec:neg} we present our results for the 
logarithmic negativity. In section~\ref{sec:two-spin} we focus on the negativity 
between two sites embedded in an infinite system, whereas in 
section~\ref{sec:half} we discuss 
the negativity between two extended regions. In section~\ref{sec:neg-spect} 
we investigate the single-particle entanglement spectrum and 
the negativity spectrum. Finally, we conclude in section~\ref{sec:concl}.

\section{Entanglement entropies, mutual information \& logarithmic negativity: Definitions}
\label{sec:obs}

In this section we introduce the definitions of the entanglement and R\'enyi 
entropies, the mutual information, and the so-called logarithmic negativity. 
Here, we always consider a system on a three-dimensional cubic lattice 
of linear size $L$ (see Fig.~\ref{fig:cartoon} (a)), prepared in 
a thermal state at temperature $T\equiv1/\beta$. 

In order to define the entanglement entropy (von Neumann entropy) and the 
R\'enyi entropies, 
we consider a bipartition of the system into two parts $A$ and $B$ 
(see Fig.~\ref{fig:cartoon} 
(b)). From the reduced density matrix for $A$, $\rho_A\equiv\textrm{Tr}_B\rho$, 
with $\rho$ being the density matrix of the full system, we define the R\'enyi 
entropies $S_{n,A}$ as~\cite{area,amico-2008,calabrese-2009,laflorencie-2016} 
\begin{equation}
\label{renyi-def}
S_{n,A}=\frac{1}{1-n}\textrm{Tr}\rho_A^n,\quad\textrm{with}\, n\in\mathbb{R}. 
\end{equation}
In the limit $n\to 1$ one obtains the von Neumann entropy as 
\begin{equation}
\label{s-vn}
S_\mathrm{vN}=-\textrm{Tr}\rho_A\ln\rho_A. 
\end{equation}
Both, the von Neumann and the R\'enyi entropies, are good entanglement 
measures for pure systems, for instance bipartite systems at zero temperature. 
This is not the case anymore if $A\cup\bar A$, with $\bar A$  the complement of $A$, 
is in a mixed state, for instance a thermal one. In 
a thermal state, $S_{n,A}$ exhibits a volume-law behaviour 
$S_{n,A}\propto L^{d}$. For $S_\mathrm{vN}$, the prefactor of the volume-law 
is the same as that of the {\it thermal} entropy, i.e., the von Neumann 
entanglement entropy becomes the thermodynamic entropy at finite 
temperature~\cite{dls-13,spr-12,kaufman-2016}. 

The mutual information between two subsystems is defined for the 
typical geometry depicted in Fig.~\ref{fig:cartoon} (c).  
The subsystem $A$ is divided into two complementary parts viz $A_1\cup A_2 = A$, with
$\bar A$ in general not empty. The mutual information $I_{n,A_1:A_2}$ is 
defined as 
\begin{equation}
\label{mi-def}
I_{n,A_1:A_2}=S_{n,A_1}+S_{n,A_2}-S_{n,A_1\cup A_2}. 
\end{equation}
Here $S_{n,X}$ is the R\'enyi entropy of subsystem $X$. 
It is important to note that the mutual information is 
only a measure of the total correlation shared between
the two subsystems, and not of the shared entanglement. 
This is because $I_{n,A_1:A_2}$ is 
sensitive to both, quantum and classical correlations, 
which is reflected in the fact that $A$ is in a mixed state. 
Importantly, this is also the case  at $T=0$ if $A_1\cup A_2$ 
is not the full system, i.e., if $A_1$ and $A_2$ are not 
complementary intervals. 
In this case the classical correlation  originates 
from the trace over the environment, i.e., $\bar A$ in~\eqref{mi-def}. 
It can be shown, however, that the mutual 
information provides a bound to the mutual entanglement~\cite{vidal-2002}.

Both the entanglement entropy and the mutual information have 
been studied intensely as possible witnesses of critical behaviour in several 
systems (see Ref.~\onlinecite{laflorencie-2016} for a review), both 
at finite-temperature phase transitions~\cite{wilms-2012,singh-2011,melko-2010,inglis-2013,jm-2014,mandal-2016,srilu-2018}, 
or at $T=0$ phase transitions~\cite{metlitski-2009,singh-2012,kallin-2013,helmes-2014,frerot-2016}. 
Specifically, in Ref.~\onlinecite{melko-2010} (see also~\onlinecite{singh-2011}) 
it has been shown that the mutual information densities $I_{n,A_1:A_2}/L^{d-1}$ 
exhibit a universal crossing at $T=T_c$ and $T=nT_c$. 
On the other hand, at zero temperature, it has been shown that the area-law 
coefficient of the entanglement entropy 
displays a cusp-like singularity across a second order phase transition (see for 
instance Ref.~\onlinecite{helmes-2014}).

Here we also consider the so-called logarithmic 
negativity~\cite{peres-1996,zycz-1998,zycz-1999,lee-2000,vidal-2002,plenio-2005}. 
Unlike the mutual information, the negativity is a good entanglement measure 
for mixed states. The negativity allows, for instance, to quantify the 
entanglement in a bipartite system 
at finite temperature (see Fig.~\ref{fig:cartoon} (b)), or the entanglement
between two non-complementary subsystems (as in Fig.~\ref{fig:cartoon} (c)) 
at zero temperature. 
The negativity is defined from the so-called partial transpose. 
Given the partition of $A$ as $A=A_1\cup A_2$ (see Fig.~\ref{fig:cartoon} (c)), 
the matrix elements of the partial transpose $\rho_A^{T_2}$ with 
respect to the degrees of freedom of $A_2$ are defined as 
\begin{equation}
\label{p-tra}
\langle\varphi_1\varphi_2|\rho_A^{T_2}|\varphi_1'\varphi_2'
\rangle=\langle\varphi_1\varphi_2'|\rho_A|\varphi_1'\varphi_2\rangle. 
\end{equation}
Here $\{\varphi_1\}$ and $\{\varphi_2\}$ are two orthonormal bases for  
$A_1$ and $A_2$, respectively. Interestingly, the eigenvalues 
$\zeta_i$ of $\rho_A^{T_2}$ can be both positive and negative. This is 
in contrast with the eigenvalues of $\rho_A$, which can be only positive. 
The logarithmic negativity ${\cal E}_{A_1:A_2}$ is defined as 
\begin{equation}
	\label{neg-def-1}
	{\cal E}_{A_1:A_2}=\ln\textrm{Tr}|\rho_A^{T_2}|. 
\end{equation}
Recently, the negativity has become a useful tool to characterise universal aspects of 
quantum many-body systems~\cite{hannu-2008,mrpr-09,hannu-2010,calabrese-2012,cct-neg-long, 
calabrese-2013,calabrese-2015,kpp-14,ruggiero-2016,rr-15,fournier-2015,bc-16,
lee-2013,castelnovo-2013,hart-2018,java-2018,bayat-2012,bayat-2014,abab-16,wen-2016,wen-2016a,ruggiero-2016a,glen,alba-e,gs-18,dct-15,gbbs-17,csg-18,sherman-2016,lu-2018,hassan-2019,xhek-2019}, 
also because it can be computed with Matrix Product States (MPS) 
methods~\cite{hannu-2008,calabrese-2013,ruggiero-2016}. 
The negativity~\eqref{neg-def-1} can be obtained for free-bosonic 
models in arbitrary dimension by correlation matrix 
techniques~\cite{audenaert-2002,eisler-2016,dct-16}. This is not possible 
for free-fermion models~\cite{eisler-2014a,coser-2015,ctc-16,chang-2016,hw-16,shapourian-2016,ssr-16,eez-16}. An alternative entanglement measure, which is effectively 
calculable using  free-fermion techniques, has been 
introduced ~\cite{shapourian-2016,ssr-16,shiozaki-2017,shiozaki-2018,shapourian-2018,shapourian-2018a}, and it is also an upper bound for the negativity ~\cite{eez-16}.  
Very recently, much attention has been focused to study the behaviour of the 
negativity at a finite-temperature phase transition. It has been suggested 
in Ref.~\onlinecite{tarun-1} and Ref.~\onlinecite{tarun-2} that in some cases the 
logarithmic negativity exhibits a cusp-like singularity, i.e., it 
is sensitive to the classical criticality, although it is a measure 
of quantum entanglement.

\section{Quantum Spherical Model}
\label{sec:model}

The quantum spherical model~\cite{HenkelHoeger, obermair-1972} (QSM) is defined on a $3D$ cubic 
lattice of volume $V=L^3$, with $L$ the lattice linear size (see Fig.~\ref{fig:cartoon}). 
The Hamiltonian  reads
\begin{equation}
\label{ham}
H = \frac{g}{2}\sum_n p_n^2 - \sum_{\langle n,m\rangle} s_ns_m + (\mu-3)\sum_n s_n^2. 
\end{equation}
In Eq.~\eqref{ham}, $n=(n_x,n_y,n_z)$ denotes a generic lattice site, and 
$\langle n,m\rangle$ a lattice bond joining two nearest-neighbour sites. 
Here $s_i$  and $p_i$ are canonically conjugated 
variables satisfying the commutation relations 
\begin{equation}
	[p_i,p_j]=[s_i,s_j]=0,\quad[s_i,p_j]=i\delta_{ij}. 
\end{equation}
In Eq.~\eqref{ham}, $\mu$ and $g$ are real parameters. 
The first term in Eq.~\eqref{ham} is a kinetic term, 
which makes the model quantum. The parameter $g$ is the quantum coupling. 
For $g=0$ the model becomes classical and 
it reduces to the famous spherical model~\cite{berlin-1952,lewiswannier}. 
The spherical parameter $\mu$ is fixed by imposing the spherical 
constraint as 
\begin{equation}
\label{constr-mu}
\sum_n \langle s_n^2\rangle=V,  
\end{equation}
where $\langle\cdot\rangle$ denotes the average over the thermal ensemble. 
The additive shift by $3$ in the definition of the spherical parameter in~\eqref{ham}
is solely for later convenience and routinely not performed in literature. 
Critical properties (for instance, critical exponents) of the QSM are 
determined by the behaviour of $\mu$. 

In order to diagonalise $H$, one exploits the translational invariance 
of the system by defining the Fourier transformed operators  $\pi_k$ and 
$q_k$ as 
\begin{equation}
	p_n = \frac{1}{\sqrt{V}}
	\sum_k e^{-i n k}\pi_k\ , \qquad s_n = \frac{1}{\sqrt{V}}
	\sum_k e^{i n k}q_k. 
\end{equation}
Here the sum over $k\equiv (k_x,k_y,k_z)$ runs in the first Brillouin zone 
$k_i\equiv \pi/L j$, with $j\in[-L,L]$ integer. In terms of $q_k,\pi_k$ 
the Fourier representation of the Hamiltonian becomes  
\begin{equation}
\label{ham-k}
 H = \sum_k \frac{g}{2}\pi_k \pi_{-k} + \Lambda_k^2 q_kq_{-k}. 
\end{equation}
The single-particle dispersion $\Lambda_k$ reads 
\begin{equation}
\label{disp}
	\Lambda_k = \sqrt{\mu + \omega_k}\quad \textrm{with} \quad 
	\omega_k = \sum_{j=x,y,z} (1-\cos k_j)
\end{equation}
To completely diagonalise~\eqref{ham-k} we introduce ladder operators $b_k$ 
and $b_k^\dagger$ as 
\begin{equation}
\label{eq:quantisation}
q_k = \alpha_k \frac{b_k+b_{-k}^\dagger}{\sqrt{2}} \ , 
\qquad \pi_k = \frac{i}{\alpha_k}\frac{b_k^\dagger - b_{-k}}{\sqrt{2}},
\end{equation}
with $\alpha_k^2 = \sqrt{g/2}\Lambda_k^{-1}$. The operators $b_k$ obey 
standard bosonic commutation relations. By using~\eqref{eq:quantisation}, 
the Hamiltonian~\eqref{ham-k} becomes diagonal, and it is given as 
\begin{equation}
\label{ham-diag}
H = \sum_k E_k (b_k^\dagger b_k + 1/2),\quad\textrm{with} \qquad E_k = \sqrt{2g} \Lambda_k
\end{equation}
In an equilibrium thermal ensemble, the modes $k$ are occupied according to the 
Bose-Einstein distribution
\begin{equation}
\label{eq:bose}
 \langle b_k b_{k'} \rangle=\langle b^\dagger_k b^\dagger_{k'} \rangle = 0 \quad, \quad
 \langle b^\dagger_k b_{k'} \rangle = \frac{\delta_{kk'}}{1-e^{\beta E_k}}, 
\end{equation}
where $\beta=1/T$ is the inverse temperature. 

\begin{figure}[t]
\includegraphics[width=0.5\textwidth]{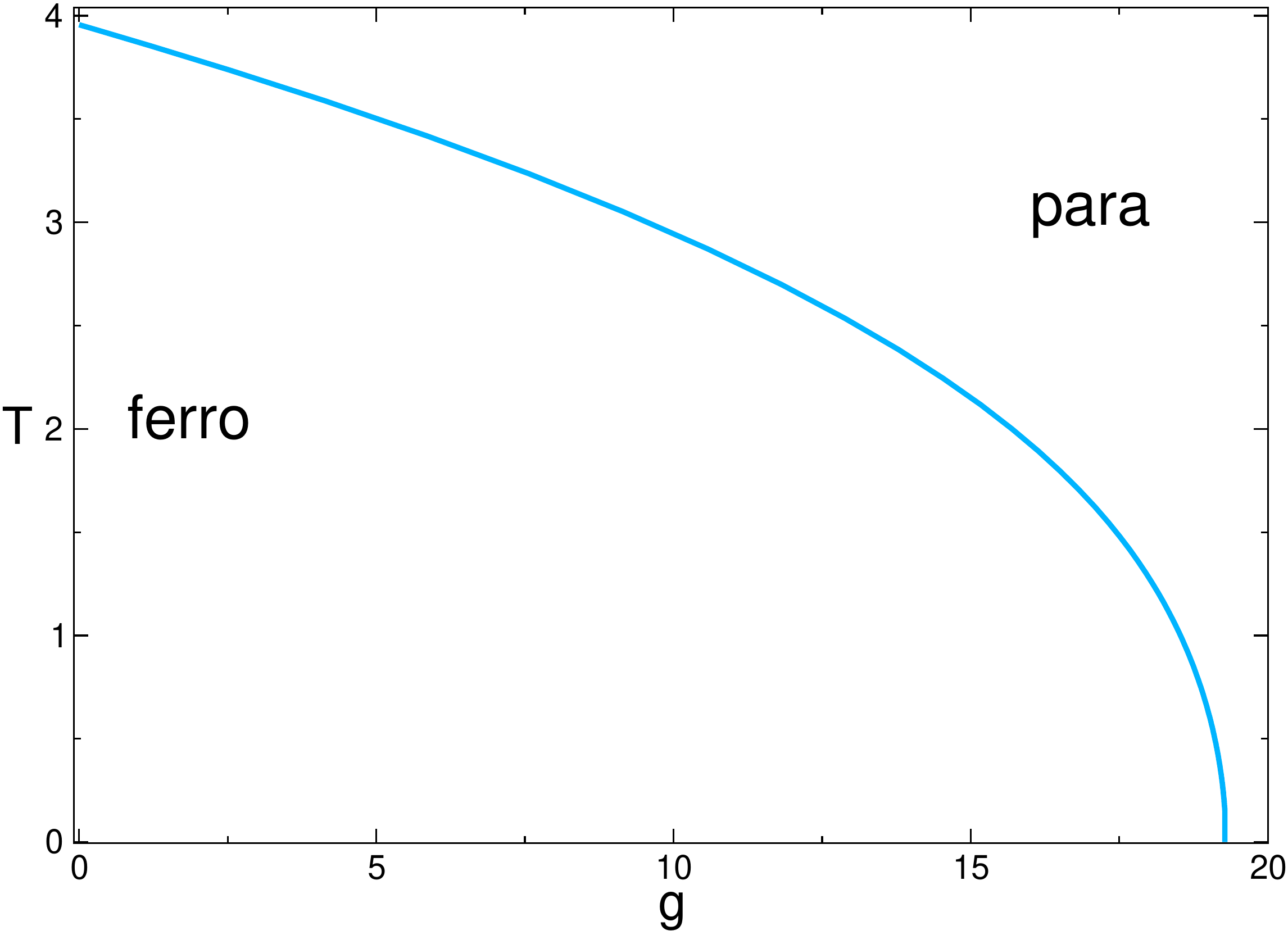}
\caption{ Phase diagram of the three-dimensional quantum spherical model. 
 $T$ is the temperature and $g$ measures the strength of quantum fluctuations (quantum coupling). 
 At low temperature and small $g$ the model is in a ferromagnetic ordered phase. This is 
 destroyed upon increasing $T$ and $g$, and the model becomes paramagnetic. The 
 continuous line marks a second-order phase transition. At $T=0$ the system undergoes 
 a quantum phase transition. At any $T\ne0$ the critical behaviour of the system 
 is in the same universality class as that of the classical spherical model at $g=0$. 
}
\label{fig:phadia}
\end{figure}

The quantum spherical model has been studied extensively in any dimension. 
In three dimensions, the model exhibits a finite-temperature 
second-order phase transition between a high-temperature paramagnetic phase and 
a low-temperature ferromagnetic (ordered) phase. The universality class of the transition 
is the same as that of the classical spherical model, and it has 
been fully characterised~\cite{joyce-1972} (see also~\cite{brankov}). The 
universality class is the same as that of the $N$-vector model at $N\to\infty$. 
At $T=0$ the model undergoes a second-order quantum phase transition at a critical  
$g_c$. The universality class of the transition is the same as that of the 
classical spherical model in $3+1$ dimensions~\cite{vojta}, as expected from 
renormalisation group arguments. 
Critical properties of the model are determined by the behaviour of 
$\mu$. For any temperature, $\mu\ge0$ for $g>g_c$. On the other hand, 
one has $\mu=0$ for $g<g_c$, i.e., in the ordered phase, which 
signals non-analytic behaviour. 
The phase diagram of the model is reported in Fig.~\ref{fig:phadia}. 
The continuous line is the critical line marking the second-order phase 
transition between the ferromagnetic phase at small $g$ and low temperature, 
and the standard paramagnetic phase.

\subsection{Two-point correlation functions}
\label{sec-qsm-corr}

The key ingredients to study entanglement-related quantities in the 
quantum spherical model are the two-point correlators of the operators 
$s_n$ and $p_n$ (cf.~\eqref{ham})
at equilibrium. They can be readily obtained by first expressing $s_n,p_n$ in terms of 
$b_k,b_k^\dagger$ that diagonalise the model, and using~\eqref{eq:bose}. One 
obtains~\cite{sascha} 
\begin{align}
\label{snsm}
\langle s_n s_m \rangle &= \frac{1}{2V}\sum_k 
e^{i (n-m)\cdot k}  \alpha_k^2 \coth (\beta E_k / 2)\\
\label{pnpm}
\langle p_n p_m \rangle &= \frac{1}{2V}\sum_k 
e^{i (n-m)\cdot k}  \alpha_k^{-2} \coth (\beta E_k / 2)\\
\label{snpm}
\langle s_n p_m\rangle &= \frac{i}{2}\delta_{nm}
\end{align}
Here we defined $k\equiv(k_x,k_y,k_z)$. 
Note that in the thermodynamic limit, at the critical point one has that 
$\mu=0$, and the contribution of 
the zero mode with $k_x=k_y=k_z=0$ diverges. This is not the case 
at finite $L$, because for a finite system $\mu$ is nonzero for 
any $g$ and $\beta$ (see section~\ref{sec:mu}). 
Clearly,  one can rewrite~\eqref{snsm} as 
\begin{equation}
\label{snsm-f}
 \langle s_n s_m \rangle 
=\frac{1}{\sqrt{V}}\mathcal{F}_d^{-1}\left(\alpha_k^2 \coth (\beta E_k / 2)\right)(n-m), 
 \end{equation}
where ${\mathcal F}^{-1}_d(x)$ denotes the inverse Fourier transform 
of $x$ in $d$ dimensions. Eq.~\eqref{snsm-f} is more suitable than~\eqref{snsm} 
for numerical computations because there are very efficient methods for 
evaluating the Fourier transform. 

From~\eqref{snsm}, the constraint~\eqref{constr-mu} for 
the spherical parameter $\mu$ reads 
\begin{equation}
\label{mu-fs}
\frac{2}{g} = \frac{1}{V}\sum_k\frac{\coth(\beta E_k / 2)}{E_k}. 
\end{equation}
The free energy $F$ of the QSM reads as 
\begin{equation}
\label{free-e}
F=TL^3\ln 2+T\sum\limits_{k_x,k_y,k_z=0}^{L-1}\ln\sinh\Big[
\frac{1}{T}\sqrt{\frac{g}{2}}\Big(\mu+3-\sum_{j=x,y,z}\cos\Big(\frac{2\pi k_j}{L}
\Big)\Big)^\frac{1}{2}\Big].
\end{equation}
The expressions for~\eqref{snsm}\eqref{pnpm}\eqref{snpm} and~\eqref{free-e} 
in the thermodynamic limit $L\to\infty$  are obtained, as usual, by replacing 
\begin{align}
	\frac{2\pi k_j}{L}&\to k'_j,\\
\frac{1}{L^3}\sum_{k_x,k_y,k_z}&\to\prod_{j=x,y,z}\int_{-\pi}^\pi \frac{dk'_j}{2\pi}. 
\end{align}
%

In the following sections  we discuss the 
behaviour of the correlators~\eqref{snsm}\eqref{pnpm}\eqref{snpm} in the 
paramagnetic phase at large $g$, and in the low-temperature ordered phase. 

\subsection{Paramagnetic phase: Large $g$ expansion}
\label{sec:large-g}

Here we discuss the large $g$ expansion of the correlators in the paramagnetic phase, 
in the thermodynamic limit. 
To this purpose, one has to first determine the behaviour of the spherical 
parameter $\mu$, which is obtained by solving Eq.~\eqref{mu-td}, in the 
large $g$ limit. We numerically verified that $\mu\to\infty$ for 
$g\to\infty$. Moreover, by taking the limit $g\to\infty$ in~\eqref{mu-td}, 
and using the definition of $E_k$ 
(cf.~\eqref{ham-k}), one obtains that the leading behaviour of 
$\mu$ is $\mu+3=g/8+o(g)$. To derive the higher order corrections, it is 
natural to conjecture that $\mu$ has the following expansion 
\begin{equation}
\label{mu-g-exp}
\mu+3=\frac{g}{8}+\frac{C_1}{g}+\frac{C_3}{g^3}+{\mathcal O}(g^{-5}).
\end{equation}
The coefficients $C_1$ and $C_3$ can be determined by substituting the ansatz 
\eqref{mu-g-exp} in~\eqref{mu-td}, expanding for $g\to\infty$, and equating 
the terms with the same power of $g$. After doing that, and after neglecting 
exponentially suppressed terms, one obtains 
\begin{equation}
C_1=9,\quad C_3=441. 
\end{equation}
Higher order terms can be obtained in a similar way. 
The behaviour of the correlation functions~\eqref{snsm} and~\eqref{pnpm}  at 
large $g$ is easily obtained after substituting~\eqref{mu-g-exp}, and 
expanding at large $g$. This yields 
\begin{align}
\label{large-g-1}
\langle s_n s_m\rangle &= \frac{\sqrt{g}}{2\sqrt{2}}
\int\frac{dk}{(2\pi)^3}
\frac{e^{i(n-m)k}}{\sqrt{\mu+\omega_k}}=
\delta_{nm}+\frac{2}{g}\delta_{|n-m|,1}+o(1/g)\\
\label{large-g-2}
\langle p_n p_m \rangle &= \frac{1}{\sqrt{2g}}
\int\frac{dk}{(2\pi)^3} 
e^{i (n-m) k}\sqrt{\mu+\omega_k}=\frac{1}{4}\delta_{nm} -\frac{1}{2g}\delta_{|n-m|,1}
+o(1/g). 
\end{align}
Here $\delta_{|n-m|,1}=\prod_{j=x,y,z}\delta_{|n_j-m_j|,1}$. 
Clearly, longer-range correlations are suppressed with higher powers of 
$1/g$, as expected because the model is paramagnetic, and correlation functions 
decay exponentially at large distances.

\subsection{Ferromagnetic phase: Low-temperature expansion}
\label{sec:small-g}

We now discuss the behaviour of the model at low temperature in the 
thermodynamic limit. In the 
ordered phase one has $\mu=0$. In the limit $\beta\to\infty$ one can 
replace $\coth(\beta x)\to 1$ in~\eqref{snsm} and~\eqref{pnpm}. 
The next-to-the-leading behaviour at large $\beta$ 
is obtained by using the standard saddle point method. One obtains the 
expansions for the correlators~\eqref{snsm} and~\eqref{pnpm} as 
\begin{align}
\label{low-exp-1}
& \langle s_ns_m\rangle=\frac{\sqrt{g}}{2\sqrt{2}}\int\frac{dk}{(2\pi)^3}\frac{e^{i k(n-m)}}
{\sqrt{\omega_k}}+\frac{1}{12\beta^2\sqrt{g}}+o(1/\beta^2)\\
\label{low-exp-2}
& \langle p_np_m\rangle=\frac{1}{\sqrt{2g}}\int\frac{dk}{(2\pi)^3}e^{i k(n-m)}
\sqrt{\omega_k}+\frac{\pi^2}{30g^{5/2}\beta^4} +o(1/\beta^4). 
\end{align}
As it is clear from~\eqref{low-exp-1} and~\eqref{low-exp-2}, the leading behaviour 
of $\langle s_ns_m\rangle$ is determined by the integral of $1/\sqrt{\omega_k}$. 
The first sub-leading correction to $\langle s_ns_m\rangle$ is 
${\mathcal O}(1/(\beta^2g^{1/2}))$, whereas for $\langle p_np_m\rangle$ it 
is ${\mathcal O}(1/(\beta^4g^{5/2}))$. Note that the sub-leading corrections 
do not depend on $n,m$. A similar behaviour occurs at criticality 
(see section~\ref{sec:critical}), and it has important consequences for 
the singularity structure of entanglement-related quantities.

\section{Critical behaviour of the quantum spherical model}
\label{sec:critical}

Here we are interested in the critical behaviour of the quantum spherical model 
at finite temperature. The main goal of this section is to derive 
the behaviour of the two-point correlators in the vicinity of the 
para-ferro transition (see Fig.~\ref{fig:phadia}). To do that we 
first derive the expansion for the spherical parameter near the 
critical point (see section~\ref{sec:mu}). In section~\ref{sec:mu-crit} 
we present our results for the correlators. In order to 
compare the behaviour of entanglement-related quantities at the finite-temperature 
phase transition with that of standard quantities,  
in section~\ref{sec:free-e} and section~\ref{sec:corr} we discuss 
the scaling of the free energy and the universal ratio $R_\xi$ 
at criticality. 

\subsection{Spherical parameter}
\label{sec:mu}

The critical behaviour of both the classical and the 
quantum spherical model is determined by the spherical parameter 
$\mu$ (cf.~\eqref{constr-mu}). It is somewhat 
easier to work in the thermodynamic limit, although the finite-size 
behaviour of $\mu$ can be derived using standard techniques~\cite{shapiro-1986}. 
In the thermodynamic limit, Eq.~\eqref{constr-mu} becomes 
(see section~\ref{sec-qsm-corr}) 
\begin{equation}
\label{mu-td}
\frac{2}{g} = \int\frac{dk}{(2\pi)^3}\frac{\coth(\beta E_k / 2)}{E_k}
\end{equation}
To proceed let us first rewrite~\eqref{mu-td} as 
\begin{equation}
\label{mu-td-1}
1=\frac{\sqrt{g}}{2\sqrt{2}}\int\frac{dk}{(2\pi)^3}
\frac{\mathrm{coth}(\beta\sqrt{g/2}\sqrt{\mu+\omega_k})}
{\sqrt{\mu+\omega_k}}, 
\end{equation}
with $\omega_k$ as defined in~\eqref{disp}.  Near criticality, 
on the paramagnetic side, one has $\mu\to0$, whereas $\mu=0$ everywhere in 
the ordered phase. This reflects the presence of singular terms in 
the expansion of $\mu$ near the transition. To derive these terms 
at the leading order in $g-g_c$, it is convenient to expand~\eqref{mu-td-1} 
for small $\mu+\omega_k$. The reason is that the singular terms 
are determined by the singularity at small $k$ of the integrand 
in~\eqref{mu-td-1}. One has 
\begin{equation}
\label{mu-td-2}
1=\frac{\sqrt{g}}{2\sqrt{2}}\int\frac{dk}{(2\pi)^3}
\frac{\mathrm{coth}(\beta\sqrt{g/2}\sqrt{\mu+\omega_k})}
{\sqrt{\mu+\omega_k}}=
\int \frac{dk}{(2\pi)^3}\Big[\frac{1}{2\beta(\mu+\omega_k)}
+a_k(\mu,\beta,g)\Big]. 
\end{equation}
Here $a(\mu,\beta,g)$ is an analytic function of its arguments. The first few 
terms of its series expansion read 
\begin{equation}
\label{mu-td-3}
	a_k(\mu,\beta,g)=\frac{g\beta}{12}-\frac{g^2\beta^3}{360}(\mu+\omega_k)+
	\frac{g^3\beta^5}{7560}(\mu+\omega_k)^2+\dots,
\end{equation}
where the dots denote higher order terms. Importantly, at the leading order in $\mu$ one 
has $a_k={\mathcal O}(\mu)$. 
The first term in~\eqref{mu-td-3} encodes the 
critical behaviour of the model. The integral is the 
celebrated Watson integral~\cite{guttmann-2010}. The same integral appears in the 
classical spherical model, reflecting that for nonzero $g$ the universality class 
of the transition is the same as that of the classical model. 
It is interesting to observe that the integral can be expressed explicitly 
in terms  of hypergeometric functions~\cite{joyce-1998}. For instance for 
$\mu=0$ one has 
\begin{equation}
\label{watson}
\int\frac{dk}{(2\pi)^3}\frac{1}{\omega_k}=\frac{\sqrt{2}}{32\sqrt{3}\pi^3}\Gamma(1/24)
\Gamma(5/24)\Gamma(7/24)\Gamma(11/24). 
\end{equation}
To proceed, one subtract from~\eqref{mu-td-3} the expansion of the equation for 
the spherical parameter~\eqref{mu-td} at $g=g_c$. One obtains  
\begin{equation}
\label{mu-td-4}
\int \frac{dk}{(2\pi)^3}\Big[-\frac{\mu}{2\beta\omega_k(\mu+\omega_k)}
+a_k(\mu,\beta,g)-a_k(0,\beta,g_c)\Big]=0. 
\end{equation}
%
\begin{figure}[t]
\includegraphics[width=0.5\textwidth]{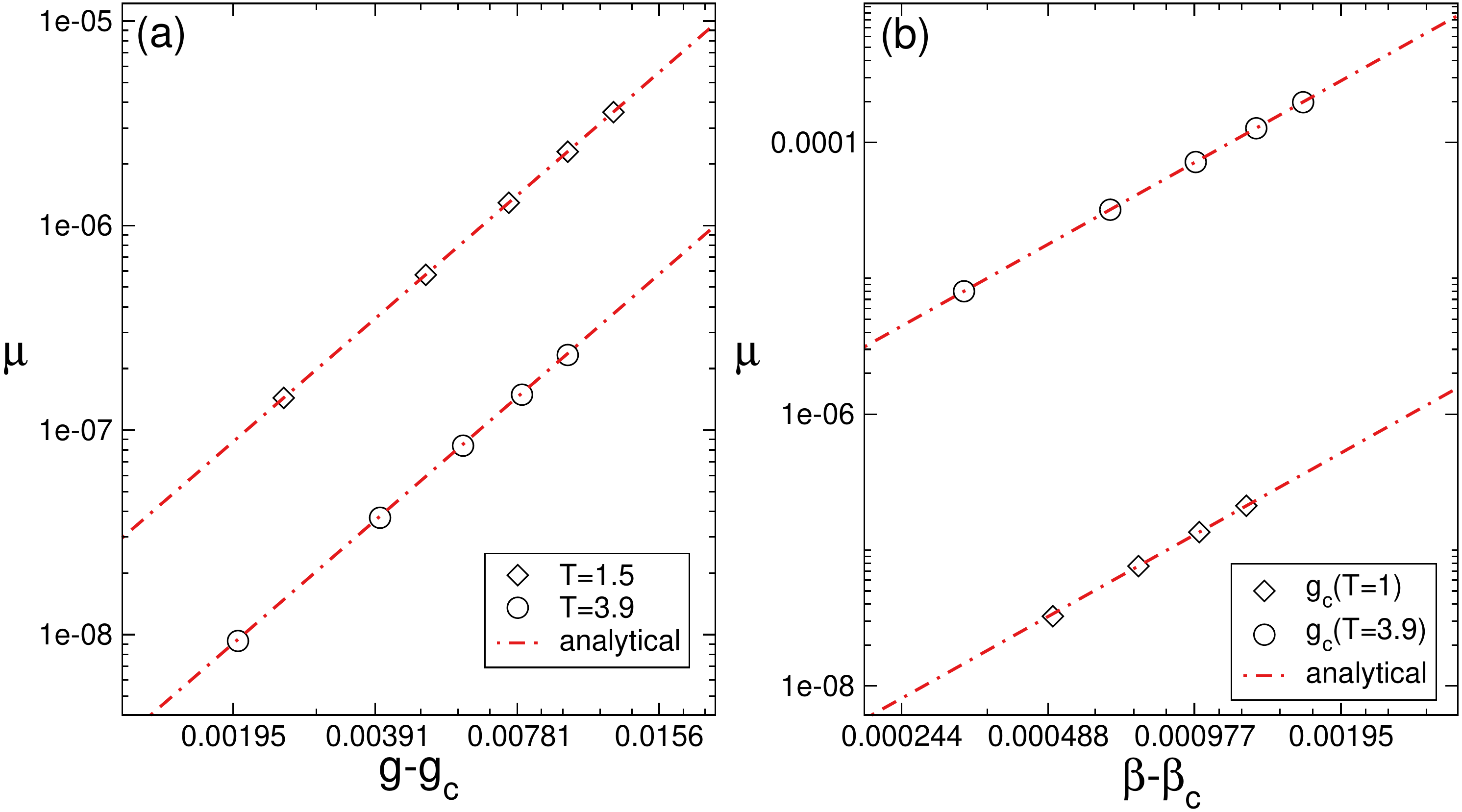}
\caption{ Spherical parameter $\mu$ near the para-ferro transition on the 
 paramagnetic side. In (a) and (b) we approach the transition at fixed temperature and fixed 
 $g$, respectively. The symbols are exact numerical results in the thermodynamic 
 limit obtained by solving~\eqref{mu-td}. The lines are the analytical results 
 near the critical point (cf.~\eqref{mu-g} and~\eqref{mu-beta}). 
}
\label{fig:mu-1}
\end{figure}

The leading behaviour in $\mu$ of the integral in~\eqref{mu-td-4} is 
obtained by expanding  $\omega_k$ for $k\to0$. One now has 
\begin{equation}
\label{mu-td-5-0}
\int \frac{dk}{(2\pi)^3}\frac{\mu}{2\beta\omega_k(\mu+\omega_k)}=
\int\frac{dk}{(2\pi)^3}\frac{\mu}{\beta|k^2|(\mu+|k|^2/2)}+{\mathcal O}(\mu)=
\frac{1}{\beta}\int_{-\pi/\sqrt{\mu}}^{\pi/\sqrt{\mu}}\frac{dy}{(2\pi)^3}
\frac{\sqrt{\mu}}{|y|^2(1+|y|^2/2)}=\frac{\sqrt{2\mu}}{4\pi\beta}+{\mathcal O}(\mu).
\end{equation}
In the last steps we changed variables to spherical coordinates with $y\equiv 
|k|$, performing the integration on a sphere of radius $\pi/\sqrt{\mu}$, instead 
of the cube $[-\pi/\sqrt{\mu},\pi/\sqrt{\mu}]$. This introduces a ${\mathcal O}(\mu)$ term. 
Since the leading order is ${\mathcal O}(\sqrt{\mu})$, this is negligible in the limit 
$g\to g_c$. Note that the term $\sqrt{\mu}$ is singular at $g_c$. 

Since we are interested in the leading ${\mathcal O}(\sqrt{\mu})$ behaviour, 
and $a_k(\mu,\beta,g)={\mathcal O}(\mu)$ we can set $\mu=0$ in $a_k$. We also observe that 
\begin{equation}
\label{mu-td-5}
a_k(0,\beta,g)-a_k(0,\beta,g_c)=
\frac{\sqrt{g}}{2\sqrt{2}}\frac{\mathrm{coth}(\beta\sqrt{g\omega_k/2})}
{\sqrt{\omega_k}}-
\frac{\sqrt{g_c}}{2\sqrt{2}}
\frac{\mathrm{coth}(\beta\sqrt{g_c\omega_k/2})}
{\sqrt{\omega_k}}. 
\end{equation}
At the leading order in $g-g_c$, from~\eqref{mu-td-3}\eqref{mu-td-4}\eqref{mu-td-5} 
we obtain 
\begin{equation}
\label{mu-g}
	\mu=\frac{\beta^2}{128\pi^4}\left[\int dk\Big(\frac{\beta}{2}+
	\frac{\mathrm{coth}(\beta\sqrt{g_c\omega_k/2})}{\sqrt{2g_c\omega_k}}-\frac{\beta}{2}
	\mathrm{coth}^2(\beta\sqrt{g_c\omega_k/2})
\Big)\right]^2(g-g_c)^2+o((g-g_c)^2).
\end{equation}
It is also convenient to expand the integrands in~\eqref{mu-g} at small $k$, to obtain 
\begin{equation}
	\mu=\frac{\pi^2\beta^4}{7200}(g-g_c)^2(20+\beta^2g_c(\beta^2g_c-4))^2+\dots
\end{equation}
In a similar way, one can derive the expression for $\mu$ if the transition is approached 
at fixed $g$ by varying the temperature. One obtains 
\begin{equation}
\label{mu-beta}
\mu=8\pi^2\left[1-\int\frac{dk}{(2\pi)^3}\Big(\frac{\beta_c g}{4}+
	\frac{\sqrt{g}\mathrm{coth}(\beta_c\sqrt{g\omega_k/2})}
		{2\sqrt{2\omega_k}}-\frac{\beta_c g}{4}
	\mathrm{coth}^2(\beta_c\sqrt{g\omega_k/2})
\Big)\right]^2(\beta-\beta_c)^2+o((\beta-\beta_c)^2).
\end{equation}
Again, after expanding for small $k$, one has 
\begin{equation}
\mu=\frac{2\pi^2}{2025}(\beta-\beta_c)^2(90+g\beta_c(-15+\beta_c^2g-6\beta_c^4g^2))^2+\dots
\end{equation}

Note that the expected behaviours~\cite{vojta} 
$\mu\propto(g-g_c)^2$ and $\mu\propto(\beta-\beta_c)^2$ hold. 
Finally, a similar calculation~\cite{shapiro-1986} 
should allow, in principle, to extract the finite-size behaviour of $\mu$. 

The correctness of~\eqref{mu-g} and~\eqref{mu-beta} is verified in Fig.~\ref{fig:mu-1}. 
The symbols in the figure are the values of $\mu$ obtained by solving~\eqref{mu-td}. 
The dashed-dotted lines are the analytical results~\eqref{mu-g} and~\eqref{mu-beta}. 
Note that~\eqref{mu-g} and \eqref{mu-beta} hold only in the vicinity of the 
transition (in the figure $g-g_c\approx10^{-2}$ and $\beta-\beta_c\approx 10^{-3}$). 

\begin{figure}[t]
\includegraphics[width=0.5\textwidth]{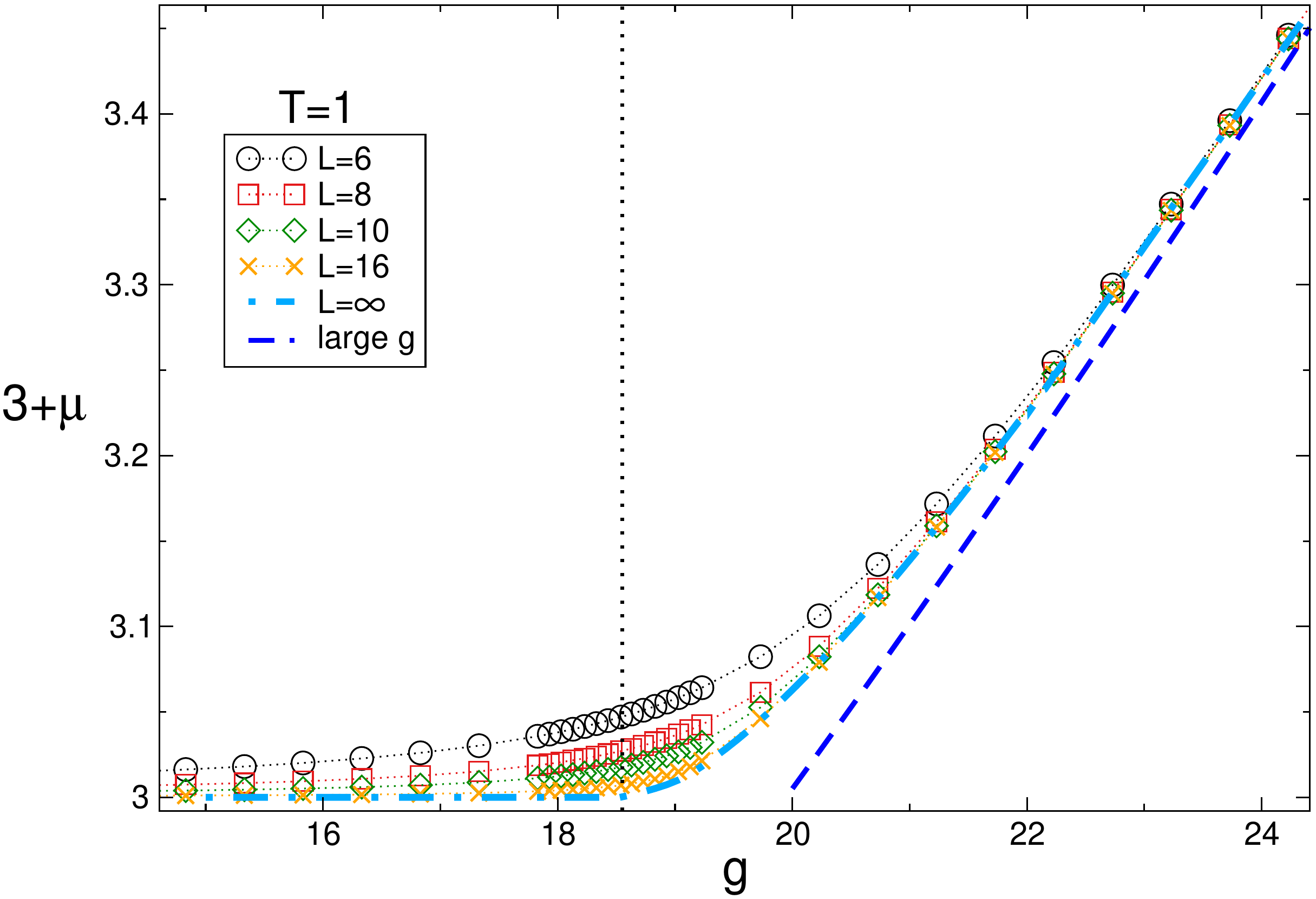}
\caption{ Spherical parameter $\mu$ in the quantum spherical model 
 at finite temperature: $3+\mu$ plotted versus $g$ at fixed temperature 
 $T=1$. The symbols are the finite-size results for several lattice sizes 
 $L$. Results are obtained by solving numerically the equation for the 
 spherical constraint~\eqref{mu-fs}. The dashed-dotted line is the 
 result in the thermodynamic limit. The dashed line is the analytical 
 result for large $g$. Note that in the thermodynamic limit 
 $\mu=0$ below the critical point at $g_c\approx 18.52$ (vertical dotted 
 line). Near the critical point it is $\mu\propto(g-g_c)^2$. 
}
\label{fig:mu}
\end{figure}

It is also useful to check how the thermodynamic limit is approached. 
In Fig.~\ref{fig:mu} we report numerical results for $\mu$ for finite-size 
systems. These are obtained by solving numerically~\eqref{mu-fs}. 
In the figure we plot $3+\mu$ versus $g$. The data are for fixed temperature 
$T=1$, although similar results 
are obtained for different temperatures. The vertical dotted line denotes 
the critical coupling $g_c$. The dashed-dotted line is the result for 
$\mu$ in the thermodynamic limit, i.e., obtained by solving~\eqref{mu-td}. 
Clearly, $\mu=0$ for $g\le g_c$, whereas $\mu\to\infty$ in the 
limit $g\to\infty$. 

The different symbols are the finite-size results for $\mu$. 
One obtains a nonzero value for $\mu$ for any $g$. As it is clear from the figure, 
both in the paramagnetic and in the ferromagnetic regions the 
data quickly converge to the thermodynamic limit (dashed-dotted line). 
Around the critical point, large finite-size effects are present. 
The perturbative result for $\mu$ (cf.~\eqref{mu-g-exp}) is reported in 
Fig.~\ref{fig:mu} as 
dashed-dotted line. Although the agreement is not perfect for the values of 
$g$ reported in the figure, we checked that the result~\eqref{mu-g-exp} is recovered 
upon increasing $g$.

\subsection{Two-point correlators}
\label{sec:mu-crit}

By using the expansion for $\mu$ near the critical line that we derived 
in the previous section, it is now straightforward to obtain the two-point correlation 
functions. Again, the idea is to expand $\omega_k$ around $k=0$ in~\eqref{snsm} and 
\eqref{pnpm} in the thermodynamic limit. For the case in which the transition is 
approached at fixed $T$, one obtains 
\begin{multline}
\label{snsm-crit-beta}
\langle s_ns_m\rangle=
\frac{\sqrt{g_c}}{2\sqrt{2}}
	\int\frac{dk}{(2\pi)^3}\frac{\coth(\beta\sqrt{g_c\omega_k/2})}{\omega_k}
	e^{ik(n-m)}\\+ 
(g-g_c)\int \frac{dk}{(2\pi)^3}\Big(\frac{\beta}{8}+
	\frac{\mathrm{coth}(\beta\sqrt{g_c\omega_k/2})}
	{4\sqrt{2g_c\omega_k}}-\frac{\beta}{8}
	\mathrm{coth}^2(\beta\sqrt{g_c\omega_k/2})
\Big)e^{ik(n-m)}-\frac{\sqrt{2\mu}}{4\pi\beta}+
{\mathcal O}((g-g_c)^2). 
\end{multline}
If the critical point is approached at fixed $g$, one has 
\begin{multline}
\label{snsm-crit-g}
\langle s_ns_m\rangle=
\frac{\sqrt{g}}{2\sqrt{2}}
	\int\frac{dk}{(2\pi)^3}\frac{\coth(\beta_c\sqrt{g\omega_k/2}}{\sqrt{\omega_k}}
	e^{ik(n-m)}+ \\
	\frac{g}{4}(\beta-\beta_c)\int \frac{dk}{(2\pi)^3}
	\Big(1-
\coth^2(\beta_c\sqrt{g\omega_k/2})
	\Big)e^{ik(n-m)}-\frac{\sqrt{2\mu}}{4\pi\beta_c}+
{\mathcal O}((\beta-\beta_c)^2). 
\end{multline}
For the correlators $\langle p_np_m\rangle$, a similar calculation yields 
\begin{multline}
\label{pnpm-crit-beta}
	\langle p_np_m\rangle=
	\frac{1}{\sqrt{2g_c}}
	\int\frac{dk}{(2\pi)^3}\coth(\beta\sqrt{g_c\omega_k/2})\sqrt{\omega_k}
	e^{ik(n-m)}+\\
	\frac{g-g_c}{g_c}\int\frac{dk}{(2\pi)^3}
	\omega_k
\Big(\frac{\beta}{4}-
	\frac{\mathrm{coth}(\beta\sqrt{g_c\omega_k/2})}{2\sqrt{2\omega_k g_c}}-
	\frac{\beta}{4}
	\mathrm{coth}^2(\beta\sqrt{g_c\omega_k/2})\Big)e^{ik(n-m)}
+{\mathcal O}((g-g_c)^2).
\end{multline}
If one approaches the transition along the temperature direction, one obtains 
\begin{multline}
\label{pnpm-crit-g}
\langle p_n p_m\rangle=
\frac{1}{\sqrt{2g}}
\int\frac{dk}{(2\pi)^3}\coth(\beta_c\sqrt{g\omega_k/2})\sqrt{\omega_k}
e^{ik(n-m)}+\\
\frac{1}{2}\int\frac{dk}{(2\pi)^3}\omega_k(1-
\coth^2(\beta_c\sqrt{g\omega_k/2}))e^{ik(n-m)}(\beta-\beta_c)+{\mathcal O}((\beta-\beta_c)^2).
\end{multline}
Clearly, in all cases (cf.~\eqref{snsm-crit-beta}\eqref{snsm-crit-g}\eqref{pnpm-crit-beta}
\eqref{pnpm-crit-g}) the correlation functions are finite at the critical point, and 
exhibit a ${\mathcal O}(g-g_c)$ and ${\mathcal O}(\beta-\beta_c)$ behaviour. 
Crucially, in both~\eqref{snsm-crit-beta} and~\eqref{snsm-crit-g} one has 
the singular contribution $\propto \sqrt{\mu}$. On the other hand, this is not present in the 
expansion of $\langle p_np_m\rangle$. This implies that the correlator 
$\langle s_ns_m\rangle$ has a cusp-like singularity across the critical point, 
whereas $\langle p_np_m\rangle$ is regular. 
Finally, one should observe that the singular term in~\eqref{snsm-crit-beta} 
and~\eqref{snsm-crit-g} does not depend on the position. This means that the 
spatial structure of the correlators appears only in sub-leading contributions 
in $g-g_c$ that we are neglecting.

\subsection{Scaling of the free energy}
\label{sec:free-e}

It is instructive to investigate the singular contributions to 
the free energy. A similar qualitative 
behaviour will be observed for entanglement-related 
quantities (see section~\ref{sec:vn}, section~\ref{sec:mi} and 
section~\ref{sec:neg}). 
Let us focus on the situation in which $T$ is kept fixed, and let us study the 
behaviour around the critical point as a function of $g$. 
At a second order phase transition the free energy {\it density} 
contains both analytic and singular terms. In general it can be 
written as~\cite{vicari} 
\begin{equation}
\label{free-rg}
f\equiv\frac{F}{L^3}= f_\mathrm{an}(g)+f_\mathrm{sing}(g). 
\end{equation}
The term $f_\mathrm{an}$ is the smooth part of the free energy, 
which depends analytically on the system parameters. The second part 
$f_\mathrm{sing}$ contains the singularities and the universal properties of 
the model. $f_\mathrm{sing}$ obeys the scaling form (see Ref.~\cite{vicari} 
for a review) 
\begin{equation}
\label{free-sc-1}
f_\mathrm{sing}(u_1,u_L,\dots,u_n)=b^{-d}f_\mathrm{sing}(b^{y_1}u_1,u_L b,\dots,b^{y_n}u_n).
\end{equation}
Here $b>0$ is an arbitrary number, and  $u_n$ are the scaling fields, 
which are analytic functions of the system's parameters. $y_n$ are  
the scaling dimensions associated with the fields $u_n$. Here we assume 
that there are only two relevant 
scaling fields, $u_1$ and $u_L$. $u_1$ is associated with changing the 
coupling $g$, and $u_L$ with the finite-size scaling. $u_L$ has scaling 
dimension $y_L=1$. We define $y_1=1/\nu$. In~\eqref{free-sc-1}, $y_n<0$ 
for $n>1$, i.e., $u_n$ are irrelevant, which give non-analytic 
scaling corrections to the free energy. Several scaling laws can be
derived from~\eqref{free-sc-1}. For instance, the finite-size 
scaling form of the free energy is obtained by choosing $b=1/u_L$. 
We obtain 
\begin{equation}
	f_\mathrm{sing}=u_L^{d}f_\mathrm{sing}(u_1/u_L^{y_1})
\end{equation}
Using that $u_L\approx 1/L$, and that close to the phase transition 
$u_1\approx g-g_c$, we find the standard result 
\begin{equation}
\label{free-fss}
f_\mathrm{sing}=L^{-d}f_\mathrm{sing}((g-g_c)L^{1/\nu}). 
\end{equation}
For a second-order phase transition, the free energy density  is finite everywhere in the 
phase diagram. The singular part $f_\mathrm{sing}$ vanishes at the critical point. 
Indeed, for $g=g_c$ one obtains from~\eqref{free-fss} the contribution 
$f_\mathrm{sing}(0)L^{-3}$, which is vanishing because $f_\mathrm{sing}$ is finite. 
By choosing $u_1b^{y_1}=1$ in~\eqref{free-sc-1} in the thermodynamic limit one obtains 
the scaling behaviour 
\begin{equation}
\label{free-e-sc-3}
f_\mathrm{sing}=|g-g_c|^{d\nu}f_\mathrm{sing}(\mathrm{sign}(g-g_c))\propto|g-g_c|^{2-\alpha}. 
\end{equation}
Note the dependence on the sign of $g-g_c$. In~\eqref{free-e-sc-3} we 
used the hyperscaling relation $d\nu=2-\alpha$, with $\alpha$ the critical exponent of the 
specific heat. In the limit $L\to\infty$ one has to recover~\eqref{free-e-sc-3} 
from~\eqref{free-fss}. This implies that $f_\mathrm{sing}((g-g_c)L^{1/\nu})\approx 
|(g-g_c)L|^{d\nu} f_\mathrm{sing}(\pm\infty)$ for $L\to\infty$. 
We should stress that in the derivations above we  neglected all the scaling corrections. 

It is interesting to derive the singular behaviour of the free energy in the 
quantum spherical model. We work in the thermodynamic limit. The density of free 
energy of the QSM reads (cf.~\eqref{free-e}) 
\begin{equation}
\label{free-e-th}
f\equiv\lim_{L\to\infty}\frac{F}{L^3}=\frac{1}{\beta}\ln(2)+\frac{1}{\beta}
\int\frac{dk}{(2\pi)^3}\ln\sinh\Big[\beta\sqrt{g/2}\sqrt{\mu+\omega_k}\Big]. 
\end{equation}
As for the spherical constraint and for the correlators (see section~\ref{sec:mu} 
and section~\ref{sec:mu-crit}), the idea is to 
expand~\eqref{free-e-th} for small $k$. After expanding~\eqref{free-e-th}, one obtains 
\begin{equation}
	\label{free-e-1}
	f=\frac{1}{\beta}(\ln(\beta\sqrt{g/2})+\ln(2))
	+\int\frac{dk}{(2\pi)^3}\Big[\frac{1}{2\beta}\ln(\mu+\omega_k)+
	b_k(\mu,\beta,g)\Big], 
\end{equation}
where $b_k$ denotes an analytic function of its arguments. 
At the leading order it is given as 
\begin{equation}
\label{bk}
b_k\equiv\frac{\beta g}{12}(\mu+\omega_k)+\dots
\end{equation}
The leading singular behaviour is encoded in the first term in~\eqref{free-e-1}. 
%
\begin{figure}[t]
\includegraphics[width=0.4\textwidth]{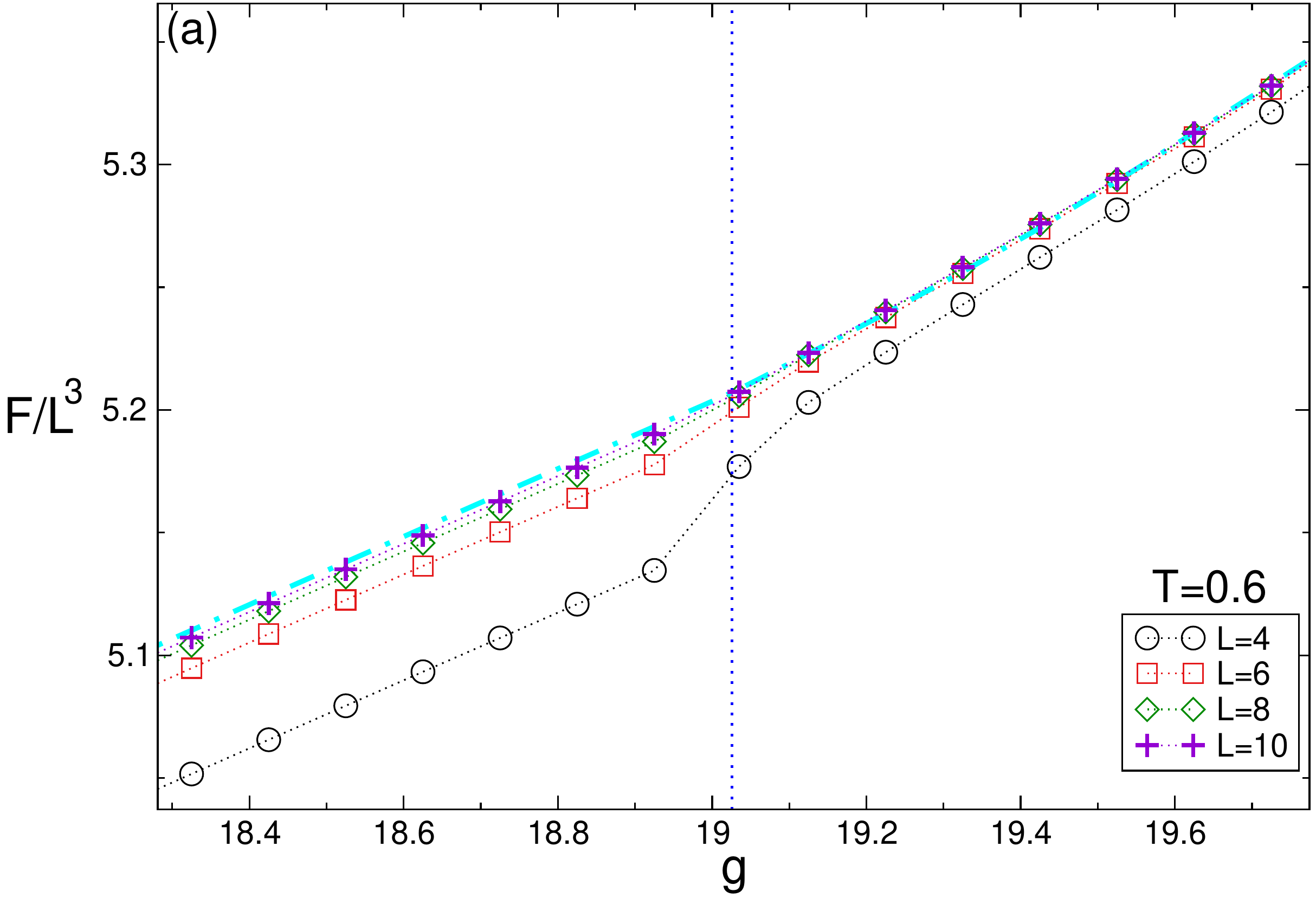}
\includegraphics[width=0.4\textwidth]{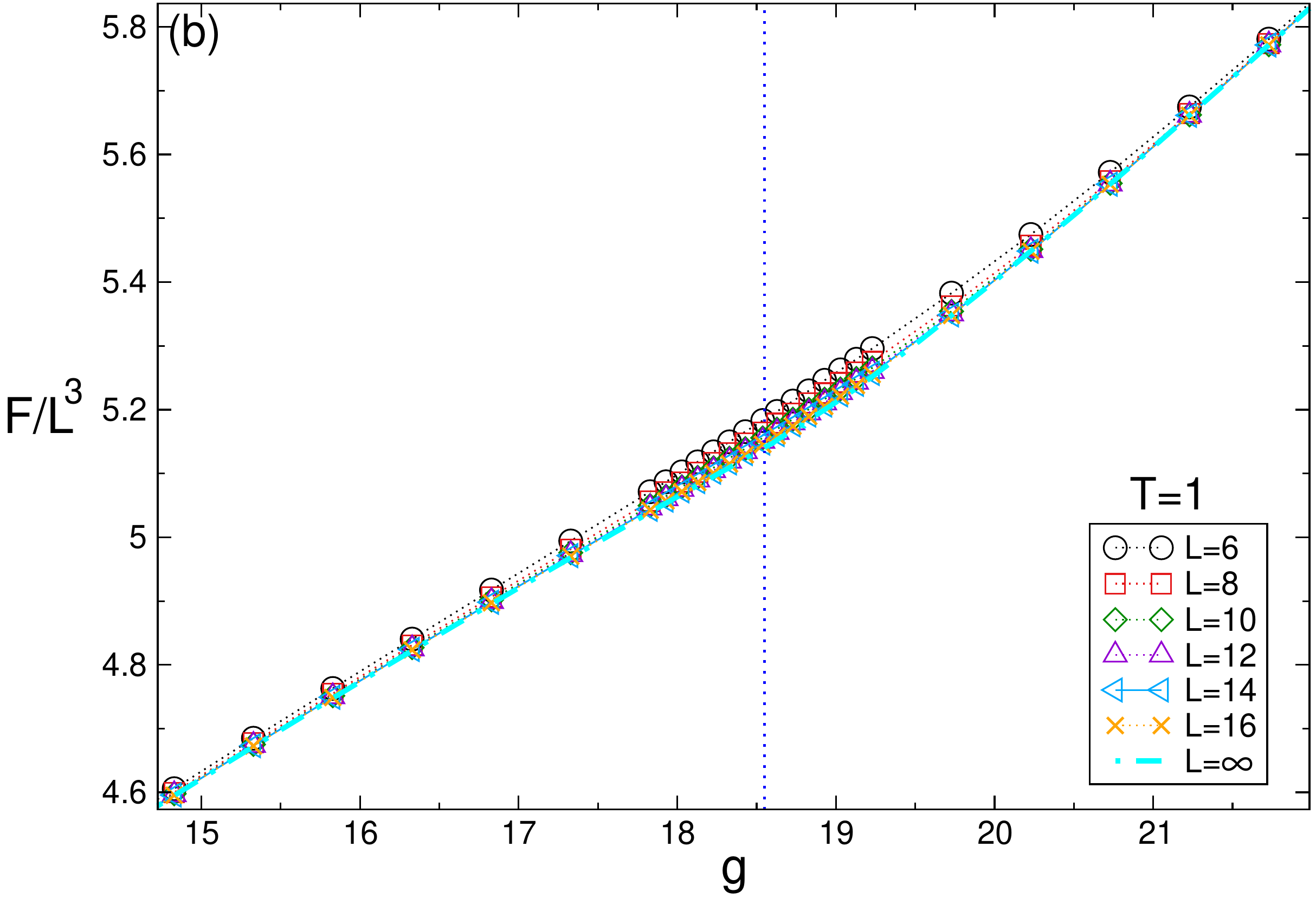}
\caption{ Density of free energy $F/L^3$ in the three-dimensional quantum spherical 
 model. (a) Results obtained by using the thermodynamic value of the spherical 
 parameter $\mu$. $F/L^3$ is plotted versus the quantum coupling $g$. Data are 
 for fixed temperature $T=0.6$ and several system sizes $L$. The singular behaviour  
 at the critical point (vertical dotted line) is unphysical (see the main text) 
 and vanish in the thermodynamic limit. The dashed-dotted line is the 
 result in the thermodynamic limit. (b) The same as in (a) using the finite-size 
 value of the spherical parameter $\mu$. Data are now for $T=1$. 
}
\label{fig:e}
\end{figure}
To extract the singularity one can use the trivial identity 
\begin{equation}
	\ln(1+z)=\int_0^\infty dt \frac{e^{-t}}{t}(1-e^{-zt}). 
\end{equation}
Now, the integration over $dk$ in~\eqref{free-e-1} can be performed exactly to give  
\begin{equation}
	\frac{1}{2\beta}\int\frac{dk}{(2\pi)^3}\ln(\mu+\omega_k)=
	\frac{1}{2\beta}\ln(3+\mu)+\frac{1}{2\beta}\int_0^\infty 
	dt (1-I^3_0(t/(3+\mu)))\frac{e^{-t}}{t}. 
\end{equation}
Here $I_0$ is the modified Bessel function of the first kind. 
Now one can split the integration range as $[0,\infty)=[0,t_0]\cup
[t_0,\infty)$. For $t_0$ large enough, by using the asymptotic behaviour of the Bessel 
function $I_0(t)\propto e^t/\sqrt{2\pi t}$, one obtains 
\begin{equation}
\label{free-int}
	\int_0^\infty dt (1-I^3_0(t/(3+\mu)))\frac{e^{-t}}{t}=\int_0^{t_0}dt
	(1-I^3_0(t/(3+\mu)))\frac{e^{-t}}{t}+\int_{t_0}^\infty dt (1-e^{3t/(3+\mu)}/(2\pi t/(3+\mu))^{3/2})\frac{e^{-t}}{t}. 
\end{equation}
The singularity of the free energy is extracted from the second term 
in~\eqref{free-int}. In the limit $\mu\to 0$ this gives 
\begin{equation}
	\label{free-int-1}
	\int_{t_0}^\infty dt (1-e^{3t/(3+\mu)}/(2\pi t/(3+\mu))^{3/2})
	\frac{e^{-t}}{t}=\Gamma(0,t_0)-\frac{\sqrt{3/2}}{(\pi t_0)^{3/2}}
	+\frac{\sqrt{3/2}(2t_0-1)}{2(\pi t_0)^{3/2}}\mu-\frac{\sqrt{2}}{3\pi}\mu^{3/2}+
	\dots
\end{equation}
Here $\Gamma(0,t_0)$ is the incomplete Gamma function. Note that both analytic and 
nonanalytic terms are present in~\eqref{free-int-1}. The term $\mu^{3/2}$ 
gives the well-known singularity of the free energy 
in the quantum spherical model~\cite{vojta} 
\begin{equation}
\label{free-e-crit}
f_\mathrm{sing}=-\frac{\sqrt{2}}{6\pi}\mu^{3/2}\propto(g-g_c)^{3},\quad
\textrm{with}\, g>g_c. 
\end{equation}
From~\eqref{free-e-crit} we obtain $\nu=1$. 
The prefactor of $(g-g_c)^3$ can be easily extracted from~\eqref{free-int-1} 
and~\eqref{mu-g}~\eqref{mu-beta}. 
Eq.~\eqref{free-e-crit} can be also derived by observing that 
the spherical constraint (see section~\ref{sec:mu}) is obtained as 
$\partial f/\partial\mu=V$, and by integrating with respect to $\mu$ 
the singular contribution in~\eqref{mu-td-5-0}. 
From~\eqref{free-e-crit}, one 
has that the specific heat exponent $\alpha=2-d\nu=-1$ is negative. 
Eq.~\eqref{free-e-crit} implies that the free energy and 
its first derivative with respect to $\beta$ and $g$ are continuous functions 
at the critical point. 

We illustrate the behaviour of the free energy across the finite temperature 
transition in Fig.~\ref{fig:e}. In the two panels (a) and (b) we 
consider two different limits. Specifically, in (a) we calculate the free energy 
density from the finite-size expression~\eqref{free-e}, but using the value of 
$\mu$ calculated in the thermodynamic limit, i.e., by solving~\eqref{mu-td}. 
This is convenient because it is straightforward to numerically 
solve~\eqref{mu-td}, whereas solving~\eqref{mu-fs} is a nontrivial task since  
the sum over $k$ cannot be performed explicitly. On the other 
hand, since in the thermodynamic limit one has $\mu=0$ below the transition, the 
contribution of the zero mode in~\eqref{free-e} is divergent and it has to 
be regularized by hand. The strategy that we use is to introduce a small 
mass putting $\mu=10^{-6}$. While this ensures that one has the correct 
thermodynamic limit results, it affects the sub-leading contributions. 
This is clear from Fig.~\ref{fig:e} (a). In the figure we show  
data for $T=0.6$ and several system sizes. The data for $L=4$ show a 
jump at $g=g_c$ (vertical line). This is an artifact of the regularization. 
We checked that the jump becomes sharper and sharper as the mass term is sent 
to zero. 
At fixed mass, upon increasing $L$, the contribution of the zero-mode becomes negligible 
and the data approach the thermodynamic limit result (dashed-dotted line in the 
figure). 
Clearly, in the thermodynamic limit the free energy and its first 
derivative are continuous, as expected from~\eqref{free-e-crit}. 
In Fig~\ref{fig:e} (b) we show the free energy 
calculated by  using the finite-size value of $\mu$, obtained by solving 
numerically~\eqref{mu-fs}. Now the contribution of the zero mode is 
finite because of the finite $L$. 

\subsection{Universal ratio $R_\xi$}
\label{sec:corr}

One important goal of this paper is to investigate the effectiveness of 
entanglement-related observables to detect finite-temperature criticality.  
In this section we  briefly review the 
behaviour of the universal ratio $\xi_\mathrm{2nd}/L$, 
with $\xi_\mathrm{2nd}$ the second-moment correlation length. This is a standard 
tool used in numerical simulations to analyse criticality~\cite{vicari}. For instance, 
it allows to detect second-order phase transitions via the so-called crossing method, 
and it can be used to extract the critical exponent $\nu$ by the usual 
data collapse analysis. 
%
\begin{figure}[t]
\includegraphics[width=0.5\textwidth]{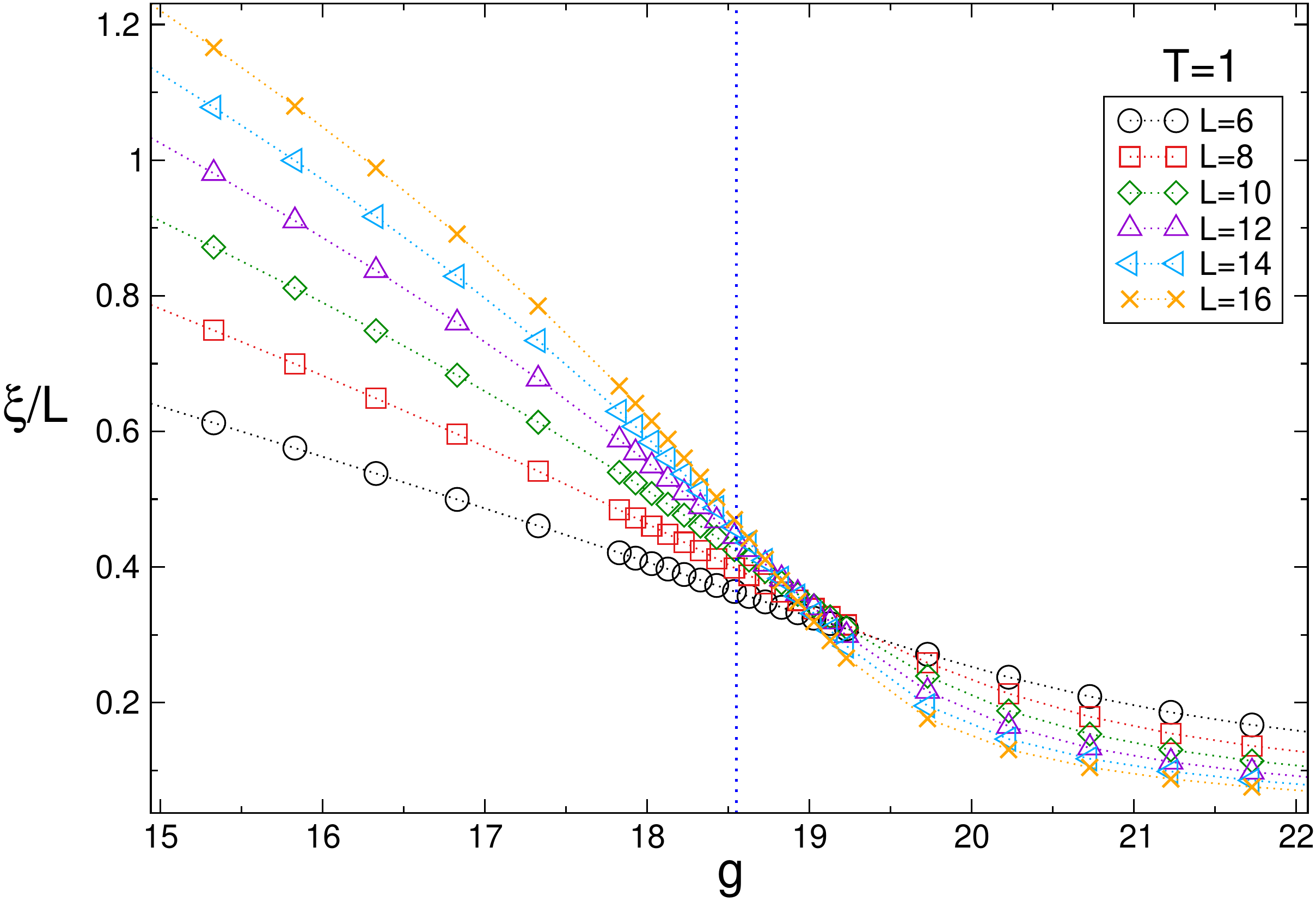}
\caption{ Scaling of the second-moment correlation length at a finite-temperature phase 
 transition in the quantum spherical model. Rescaled correlation 
 length $\xi_\mathrm{2nd}/L$, with $L$ the system size, plotted as a function of $g$. 
 The data are for fixed $T=1$ and several $L$. The vertical dotted line 
 marks the critical point at $g_c\approx 18.52$. Note the crossing at $g\approx g_c$ 
 between curves for different system sizes. 
}
\label{fig:xi}
\end{figure}
%
\begin{figure}[t]
\includegraphics[width=0.5\textwidth]{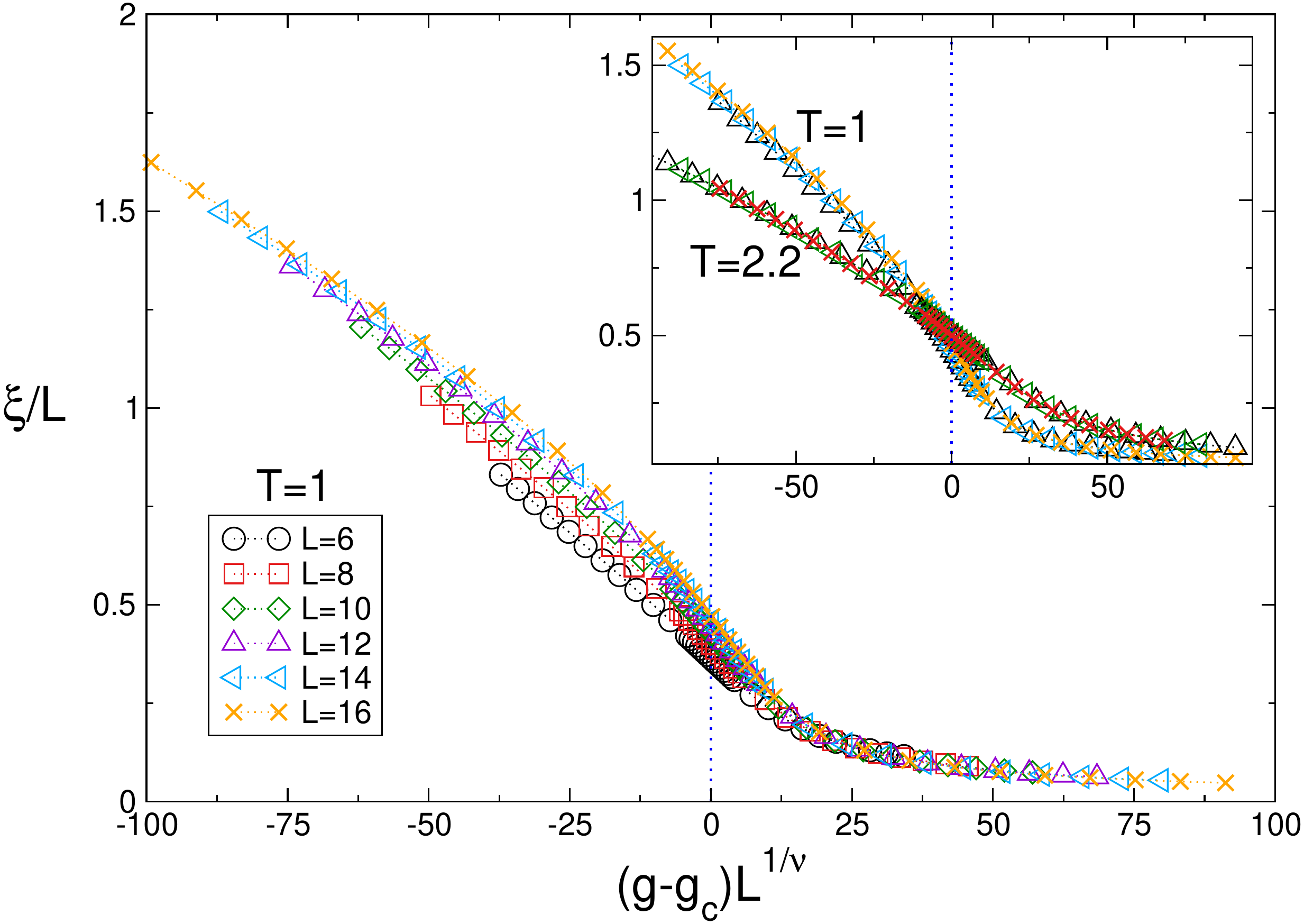}
\caption{ 
 Same data as in Fig.~\ref{fig:xi} plotted versus the scaling variable 
 $X\equiv(g-g_c)L^{1/\nu}$, with $\nu=1$. 
 Upon increasing $L$, the data for different $L$ collapse on the same 
 curve. Note the large scaling corrections near the critical 
 region at $X=0$. Inset: Zoom around $X=0$ showing only 
 data for $L=12$-$16$. We now show results also for $T=2.2$. Note 
 for both temperatures the same universal value 
 $\xi_\mathrm{2nd}/L\approx 0.5$ at the critical point $X=0$. 
}
\label{fig:xi-1}
\end{figure}
%
The ratio $R_\xi$ is defined as 
\begin{equation}
	R_\xi=\frac{\xi_\mathrm{2nd}}{L}. 
\end{equation}
The so-called second-moment correlation length 
$\xi_\mathrm{2nd}$ is extracted from the long-distance behaviour of the 
correlation function. Its definition reads 
\begin{equation}
	\label{xi-2nd}
	\xi_\mathrm{2nd}^2 = \frac{\widetilde{G}(0)/\widetilde{G}
	(q_\mathrm{min}) -1}{4\sin^2(\pi/L)}, 
\end{equation}
with $\widetilde{G}(k)$  the Fourier transform of the spin-spin correlation 
function (cf.~\eqref{snsm} for the spherical model). 
In~\eqref{xi-2nd}, $q_\mathrm{min}$ is the minimum nonzero 
lattice momentum $q_\mathrm{min}=(2\pi/L,0,0)$. 
Note that $\widetilde G(0)$ is the spin susceptibility $\chi$ 
defined as 
\begin{equation}
\label{chi}
\chi=\frac{1}{V}\sum\limits_{n,m}\langle s_ns_m\rangle. 
\end{equation}
Alternatively, instead of $\xi_\mathrm{2nd}$, one can define $R_\xi$ 
by using $\xi_\mathrm{gap}$, 
which is defined from the gap between the ground state and the first 
excited state in the energy spectrum. 
For the quantum spherical model $R_\xi$ can be expressed analytically 
as a function of $\mu$. By using~\eqref{snsm}, one obtains that the 
correlation length is given as 
\begin{equation}
\xi^2_\mathrm{2nd} = \frac{1}{4\sin^2(\pi/L)}\left[
\frac{\alpha_{2\pi/L}^2 \coth(\beta E_{2\pi/L}/2)}{\alpha_0^2 
\coth (\beta E_0 / 2)} -1\right]
\end{equation}
We now discuss the finite-size scaling properties of $R_\xi$. 
The finite-size scaling ansatz for $R_\xi$ reads as~\cite{vicari} 
\begin{equation}
\label{rxi}
R_\xi=K((g-g_c)L^{1/\nu}). 
\end{equation}
Here we neglect both analytic and nonanalytic scaling corrections, that, however, 
one can include. The function $K(x)$ is universal apart from a renormalisation 
of its argument. Assuming analyticity of $K(x)$ one can expand~\eqref{rxi} as 
\begin{equation}
\label{rxi-exp}
R_\xi=R^*_\xi +c_1(g-g_c)L^{1/\nu}+\dots. 
\end{equation}
The value $R^*_\xi$ depends only on the universality class of the 
transition and on the geometry (for instance, the boundary conditions). 
Crucially, Eq.~\eqref{rxi-exp} implies that 
the curves for $R_\xi$ at different 
finite sizes $L$ exhibit a crossing at the critical point. 
Moreover, 
when plotted against the scaling variable $X\equiv (g-g_c)L^{1/\nu}$, 
the data for different $L$s collapse on the same curve, at least in 
the limit $L\to\infty$, when corrections to scaling can be neglected.  

It is important to understand the behaviour of $R_\xi$, especially 
with the moderately small system sizes that are available in 
our simulations. In Fig.~\ref{fig:xi}
we show numerical data for $R_\xi$ plotted as a function of $g$. 
The data are for $T=1$. The curves for different $L$ exhibit a 
crossing at $g\approx g_c$, as expected. The value $R_\xi^*\approx 0.5$ at 
the crossing is universal, and it could be calculated 
by using the results of Ref.~\onlinecite{shapiro-1986}. To 
our knowledge $R_\xi^*$ is not known exactly.  
Note that scaling corrections are present. Indeed, for the 
smaller $L$s the crossing is not at $g_c$ 
(vertical line). Upon increasing $L$, the crossing point exhibits a systematic 
drifts towards $g_c$. In Fig.~\ref{fig:xi-1}  we show the same 
data as in Fig.~\ref{fig:xi}, now plotted versus the scaling variable 
$X=(g-g_c)L^{1/\nu}$. As expected, due to the scaling corrections, the 
smaller lattice sizes do not show data 
collapse. However, upon increasing $L$ the quality of the collapse improves 
significantly. The scaling is satisfactory for the larger sizes. 
In the inset we provide data also for $T=2.2$. We observe that scaling 
corrections are smaller as compared with $T=1$,  which is reflected in a better data 
collapse. Note that the scaling function $K$ (cf.~\eqref{rxi}) 
depends on the temperature, although the value at $X=0$ is the same for both 
temperatures, as expected. 
However, since the finite-temperature universality class is the same on the whole 
para-ferro transition line (see Fig.~\ref{fig:phadia}), the two scaling functions 
should coincide after an analytic redefinition of the scaling field $u_1$ as 
$u_1\approx c_T(g-g_c)$, where $c_T$ depends on the temperature. We numerically 
verified that $c_{T=2.2}/c_{T=1}\approx1/2$.

\section{Entanglement scaling in the critical spherical model}
\label{sec-ent-qsm}

We now discuss the behaviour of entanglement-motivated quantities in the 
finite-temperature critical QSM. Specifically, in 
section~\ref{sec:ent-cal} we briefly review how to calculate 
entanglement-related observables. In 
section~\ref{sec:vn} we focus on the von Neumann entropy. In 
section~\ref{sec:mi} we discuss the von Neumann and R\'enyi mutual 
information. In section~\ref{sec:neg} we investigate the 
the logarithmic negativity. We consider both the negativity between 
two spins in an infinite system (in section~\ref{sec:two-spin}), 
as well as the negativity between two extended adjacent blocks (in 
section~\ref{sec:half}). Finally, in section~\ref{sec:neg-spect} 
we focus on the single-particle entanglement spectra and 
negativity spectra. 

\subsection{Computation of entanglement-related observables in the QSM}
\label{sec:ent-cal}

The quantum spherical model is mappable to a system of free bosons 
(see section~\ref{sec:model}). This implies that entanglement-related quantities are obtained 
from the two-point correlation functions (see Ref.~\cite{viktor} for a 
review). The key ingredients are the matrices 
$\mathbb{Q}_{nm}\equiv\langle s_ns_m\rangle$ and $\mathbb{P}_{nm}\equiv\langle p_np_m
\rangle$ constructed from the two-point correlation functions 
(cf.~\eqref{snsm} and~\eqref{pnpm}). 

Let us define the matrices $\mathbb{Q}[A]$ and $\mathbb{P}[A]$, 
where $n,m$ are now restricted to subsystem $A$ (see Fig.~\ref{fig:cartoon} 
(b) and (c)). The R\'enyi entropies of $A$ are constructed from 
the eigenvalues $\lambda_j^2$ of the matrix $\mathbb{Q}[A]\cdot\mathbb{P}[A]$. 
As it is common in the literature, we refer to the $\lambda_j$ as the single-particle 
entanglement spectrum levels. In terms of $\lambda_j$, 
the R\'enyi entropies are given as 
\begin{equation}
\label{renyi-boson}
S_{n,A}=-\frac{1}{1-n}\sum_j\ln\Big[
	\Big(\lambda_j+\frac{1}{2}\Big)^n-\Big(\lambda_j-\frac{1}{2}\Big)^n\Big]. 
\end{equation}
The von Neumann entropy is obtained by performing the analytic 
continuation $n\to 1$, which yields 
\begin{equation}
\label{vN-boson}
S_\mathrm{vN}=\sum_j\Big[\Big(\lambda_j+\frac{1}{2}\Big)\ln\Big(\lambda_j+\frac{1}{2}\Big)
	-\Big(\lambda_j-\frac{1}{2}\Big)\ln\Big(\lambda_j-\frac{1}{2}\Big)\Big]. 
\end{equation}
The mutual information (see section~\ref{sec:obs}) is obtained 
by using~\eqref{renyi-boson}~\eqref{vN-boson} and~\eqref{mi-def}. 

We now discuss the logarithmic negativity for a partition of $A$ as $A=A_1\cup A_2$ 
(see Fig.~\ref{fig:cartoon} (c)). One 
first defines the transposed matrix $\mathbb{P}[A^{T_2}]$ as
\begin{equation}
	\mathbb{P}[A^{T_2}]\equiv 
	\mathbb{R}[A^{T_2}]\mathbb{P}[A^{T_2}]
	\mathbb{R}[A^{T_2}]. 
\end{equation}
Here the matrix $\mathbb{R}[A^{T_2}]$ acts as the identity matrix 
$\mathbb{I}_{A_1}$ on $A_1$ and as $-\mathbb{I}_{A_2}$ on $A_2$. 
The eigenvalues $\nu_i^2$ of $\mathbb{Q}[A]\cdot\mathbb{P}[A^{T_2}]$ 
form the single-particle negativity spectrum. In terms of $\nu_i^2$, 
the negativity is given as 
\begin{equation}
\label{neg-boson}
{\cal E}=\sum_i\textrm{max}(0,-\ln(2\nu_i)). 
\end{equation}
Note that while $\nu_i^2>0$, $-\ln(2\nu_i)$ can be both positive and 
negative. 

\subsection{Von Neumann entropy}
\label{sec:vn}

\begin{figure}[t]
\includegraphics[width=0.5\textwidth]{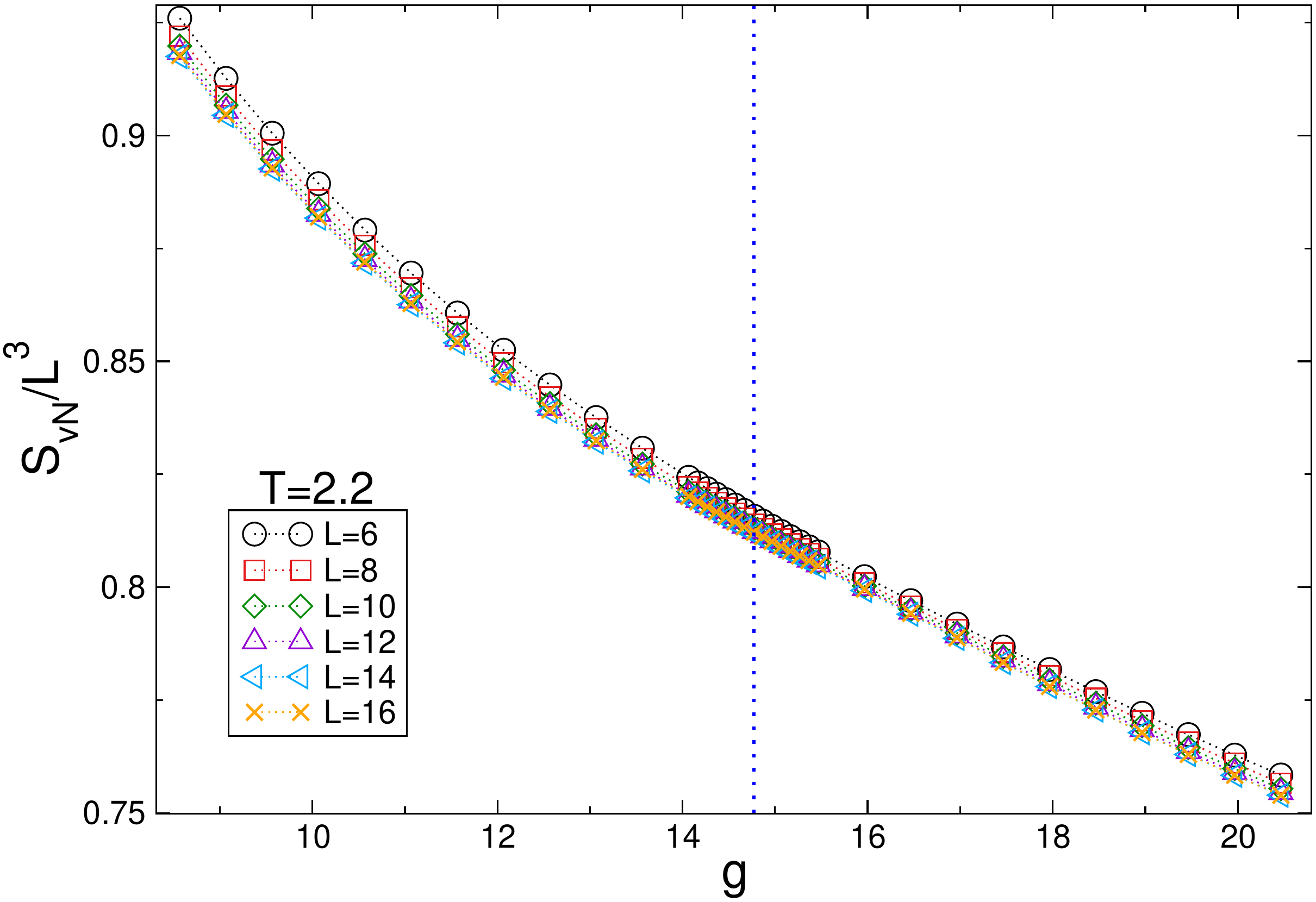}
\caption{ Scaling of the entanglement entropy density $S_\mathrm{vN}/L^3$ 
 at a finite-temperature quantum phase transition in the quantum spherical 
 model. The figure shows $S_\mathrm{vN}/L^3$ as 
 a function of $g$ for several system sizes $L$. The vertical line marks the critical 
 point. 
}
\label{fig:entropy}
\end{figure}

Let us start discussing the behaviour of the von Neumann entropy. 
Numerical data for the half-system entropy are reported in 
Fig.~\ref{fig:entropy}, for fixed temperature $T=2.2$. 
Similar to the free energy (see Fig.~\ref{fig:e}) and to the thermal entropy, 
the von Neumann entropy exhibits a volume law at any $g$. 
The figure shows the density of entropy $S_\mathrm{vN}/L^3$, plotted as  a function of 
$g$. Different symbols are for different system sizes. 
Finite-size effects decay very quickly with $L$. The data for the 
larger sizes $L=12-16$ collapse on the same curve. The entropy density exhibits 
regular behaviour around the transition. This is expected because at finite 
temperature the density of von Neumann entanglement entropy becomes the same 
as the thermal entropy, which is not singular at the transition. 
We should stress, however, that sub-leading 
contributions can be singular, for instance due to the presence of the 
zero mode. A natural scaling ansatz for the entropy density reads as 
\begin{equation}
\label{ent-sc}
\frac{S_\mathrm{vN}}{L^3}=L^{-3}s_\mathrm{sing}((g-g_c)L^{1/\nu})+s_\mathrm{an}(g). 
\end{equation}
Here the functions $s_\mathrm{sing}$ and $s_\mathrm{an}$ encode the singular and regular 
terms. 
In~\eqref{ent-sc} we neglect scaling corrections. 
Eq.~\eqref{ent-sc} implies that the 
singular term vanishes at the transition when increasing $L$, and only the 
regular term survives. We should stress that in~\eqref{ent-sc} we also neglect
logarithmic contributions that can arise because of the presence of 
the zero mode~\cite{met-grov}, which reflects the symmetry breaking. 
We will investigate these contributions in section~\ref{sec:neg-spect} discussing  
the single-particle entanglement spectrum. Moreover, in principle, there can be 
extra logarithmic corrections if the bipartition has corners (see, for instance 
Ref.~\onlinecite{singh-2012}). These are not present 
in our case because the boundary between $A$ and its complement is smooth 
(see Fig.~\ref{fig:cartoon}).

\subsection{Mutual information}
\label{sec:mi}

\begin{figure}[t]
\includegraphics[width=0.5\textwidth]{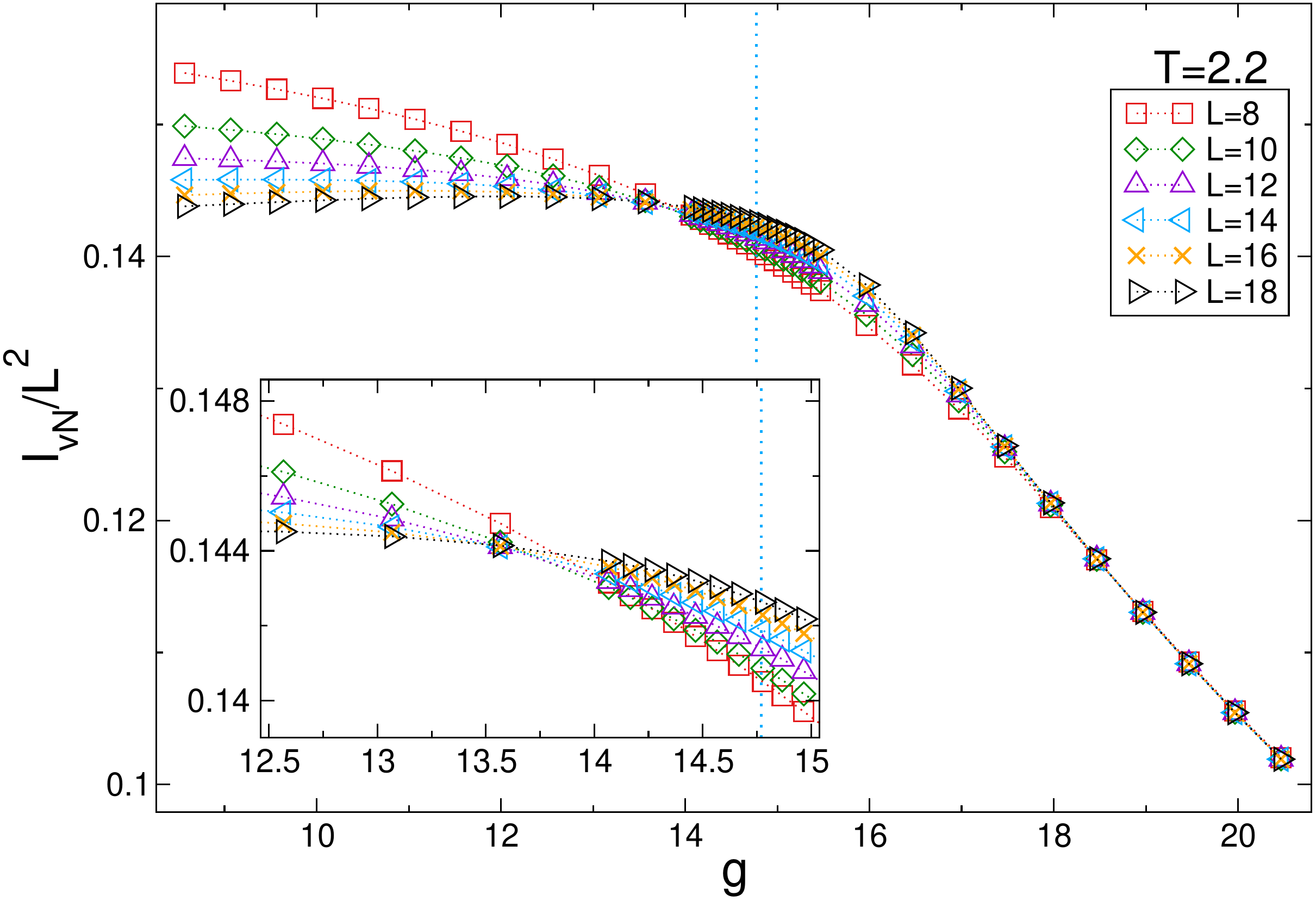}
\caption{ Scaling of the density of von Neumann mutual information $I_\mathrm{vN}/L^2$ 
 at a finite-temperature quantum phase transition in the quantum spherical 
 model: $I_\mathrm{vN}/L^2$ as  a function of $g$ for several system sizes $L$. 
 Data are for fixed $T=2.2$. The vertical dotted line marks the critical 
 point at $g_c$. Note the spurious crossing at $g<g_c$. 
}
\label{fig:mi}
\end{figure}

\begin{figure}[t]
\includegraphics[width=0.5\textwidth]{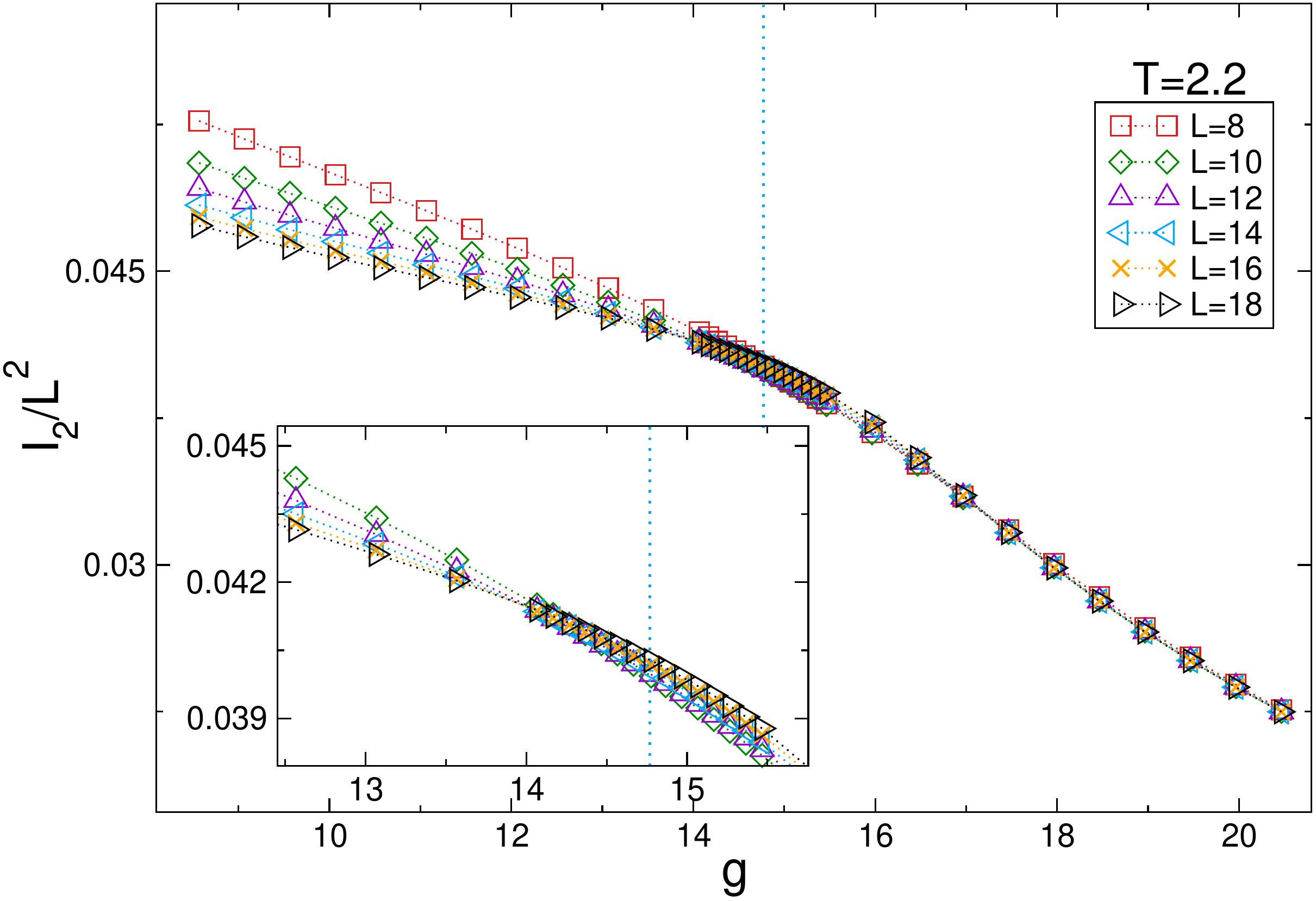}
\caption{ Scaling of the density of R\'enyi mutual information $I_{n}/L^2$ with 
 $n=2$ at a finite-temperature phase transition: $I_{n}/L^2$ as  a function 
 of $g$ for several system sizes $L$. Data are for fixed $T=2.2$. The vertical 
 dotted line marks the critical point at $g_c$. Inset: zoom around the critical 
 region. Note that no crossing is present. 
}
\label{fig:mi-n2}
\end{figure}

We now turn to the mutual information (see~\eqref{mi-def}  for 
its definition). 
Here we consider a bipartite system at finite temperature (as in Fig.~\ref{fig:cartoon} 
(b)). The system is divided into two equal parts $A$ and $B=\bar A$. We consider 
the mutual information between $A$ and its complement. 
As it is clear from~\eqref{mi-def}, the volume-law contribution of 
the entropies cancels out. This cancellation happens for any value of $g$ and $T$. 
Thus, the mutual information exhibits an area law. 
A natural scaling ansatz for the density of the mutual information is  
\begin{equation}
\label{mi-sc}
\frac{I_{n}}{L^2}=L^{-2}q^{(n)}_\mathrm{sing}((g-g_c)L^{1/\nu})+q^{(n)}_\mathrm{an}(g). 
\end{equation}
Eq.~\eqref{mi-sc} is compatible with the scaling ansatz proposed in 
Ref.~\onlinecite{singh-2011}. 
Here $\nu$ is the critical exponent as in~\eqref{ent-sc}. Similar to~\eqref{ent-sc}, 
here we are neglecting scaling corrections. In~\eqref{mi-sc}, $q_\mathrm{an}$ is 
the analytic contribution. 

We present our results for the von Neumann mutual information $I_\mathrm{vN}$ in Fig.~\ref{fig:mi}. 
We plot the density of mutual information $I_\mathrm{vN}/L^2$ versus $g$.  
Data are at fixed $T=2.2$. The data exhibit a crossing point around 
$g\approx 13.5$, which is incompatible with the critical point at $g_c\approx 14.77$. 
Moreover, the position of the crossing point does not change upon increasing system 
size, in contrast with the behaviour of $R_\xi$ (see Fig.~\ref{fig:xi}). 
Importantly, the curves for different $L$s do not ``fan out'' as $L$ increases, 
in contrast with the behaviour for the ratio $R_\xi$ (see~\ref{fig:xi}). As for the 
von Neumann entropy, the mutual information is dominated by the regular part 
(cf.~\eqref{mi-sc}). More precisely, although the singular term 
in~\eqref{mi-sc} gives a universal crossing, this is not visible because 
it is suppressed as $L^{-2}$ in the limit $L\to\infty$.

Similar behaviour is observed for the R\'enyi mutual information. This is discussed 
in Fig.~\ref{fig:mi-n2} focusing on $I_2$. 
As for $I_\mathrm{vN}$, the data in Fig.~\ref{fig:mi-n2} 
collapse on the same curve upon increasing $L$. Specifically, in the paramagnetic phase 
for $g>g_c$ and close to the critical point, finite-size effects decay 
dramatically with $L$, and the data with $L\gtrsim10$ 
are already indistinguishable from the thermodynamic limit. 
Also no universal crossing is visible within the 
system sizes presented in the Figure. 

It is interesting to compare our results with Ref.~\onlinecite{singh-2011}. 
There is large evidence, based on quantum Monte Carlo simulations, that at a 
finite-temperature phase transition the ratio $I_2/L^{d-1}$ 
exhibits two crossing at $T_c$ and $2T_c$, with $T_c$ the critical temperature. 
The scaling ansatz for the mutual information presented in Ref.~\onlinecite{singh-2011} 
is compatible with~\eqref{mi-sc}. However, as stressed in Ref.~\onlinecite{singh-2011} 
the presence of the crossing relies on $q_\mathrm{an}^{(n)}$ changing sign across the 
phase transition. Our results suggest that this does not happen in the 
quantum spherical model.

\section{Logarithmic negativity}
\label{sec:neg}

\begin{figure}[t]
\includegraphics[width=0.6\textwidth]{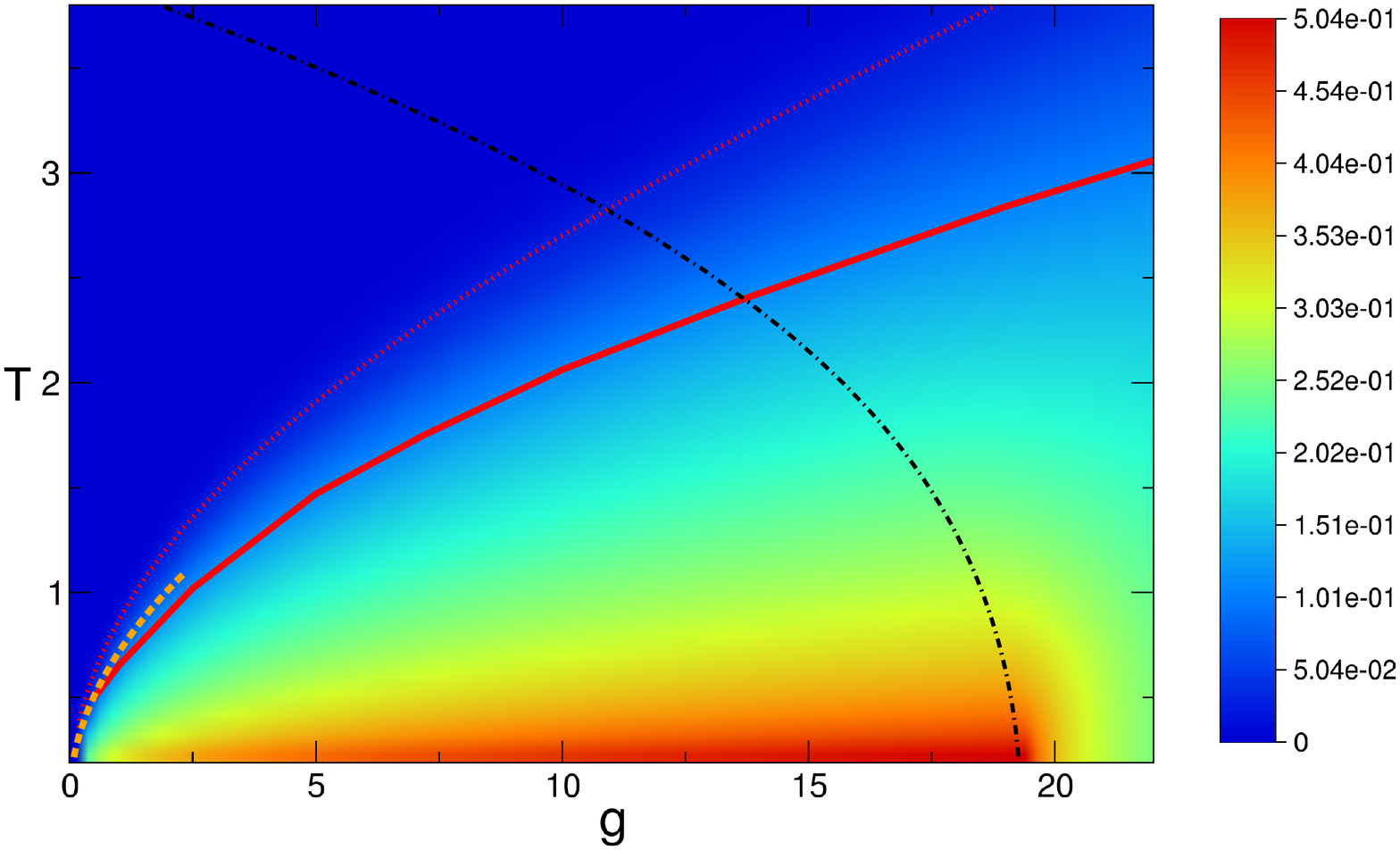}
\caption{ Survey of the behaviour  of the logarithmic negativity in the 
 quantum spherical model. On the $x$-axis $g$ is the quantum coupling. 
 On the $y$-axis $T$ is the temperature. The figure shows the density 
 plot of the half-system negativity. The system is a cube of 
 length $L=2$. The dashed-dotted line is the critical line 
 dividing the paramagnetic phase from the ordered phase at low 
 temperature. The dotted line is the ``death line'' of the negativity. 
 Above the death line the negativity is exactly zero. The continuous 
 line is the death-line calculated from the negativity between two adjacent 
 spins embedded in an infinite system. The behaviour as $\propto 
 g^{1/2}$ at small $g$ is also reported (dashed  line). 
}
\label{fig:neg_den}
\end{figure}
%

We now discuss the logarithmic negativity between two 
complementary subsystems. This allows us to study the interplay 
between genuine quantum fluctuations and thermal  
fluctuations. One of the goals of this section is to map out 
the role of entanglement in the different regions of the 
phase diagram of the quantum spherical model. 

The generic behaviour of the negativity is illustrated in Fig.~\ref{fig:neg_den}. 
The figure shows a density plot of ${\cal E}$ as a function of $g$ and 
$T$. The dashed-dotted line is  the critical line dividing the paramagnetic 
phase from the ordered phase (see Fig.~\ref{fig:phadia}). 
In the figure we show the half-system negativity for a cube of linear size $L=2$. 
The data are obtained by using the value of $\mu$ in the thermodynamic limit 
and the finite-size formulas for the correlators (cf.~\eqref{snsm}\eqref{pnpm}). 
The negativity is large at the quantum critical point and in the ordered phase, 
and it quickly decays upon increasing the temperature. The dotted line is 
the negativity ``death line''. Above the 
death line the half-system negativity is exactly zero. The death line that we report in 
the figure is obtained by considering the half-system negativity in the 
limit $L\to\infty$. In the figure we also report the death line calculated from 
the negativity between two adjacent sites embedded in an infinite 
system (continuous line), which provides only a bound. Interestingly, 
the death line exhibits the behaviour $\propto g^{1/2}$ at small $g$. For 
the case of two sites embedded in the infinite system the precise behaviour 
can be calculated analytically and it is reported in the figure (dashed line). 
Finally, it is interesting to observe that the death line increases with  
$g$, although the negativity decreases as $1/g$ (see 
section~\ref{sec:two-large-g}).

\subsection{Two-site negativity}
\label{sec:two-spin}

%
\begin{figure}[t]
\includegraphics[width=0.5\textwidth]{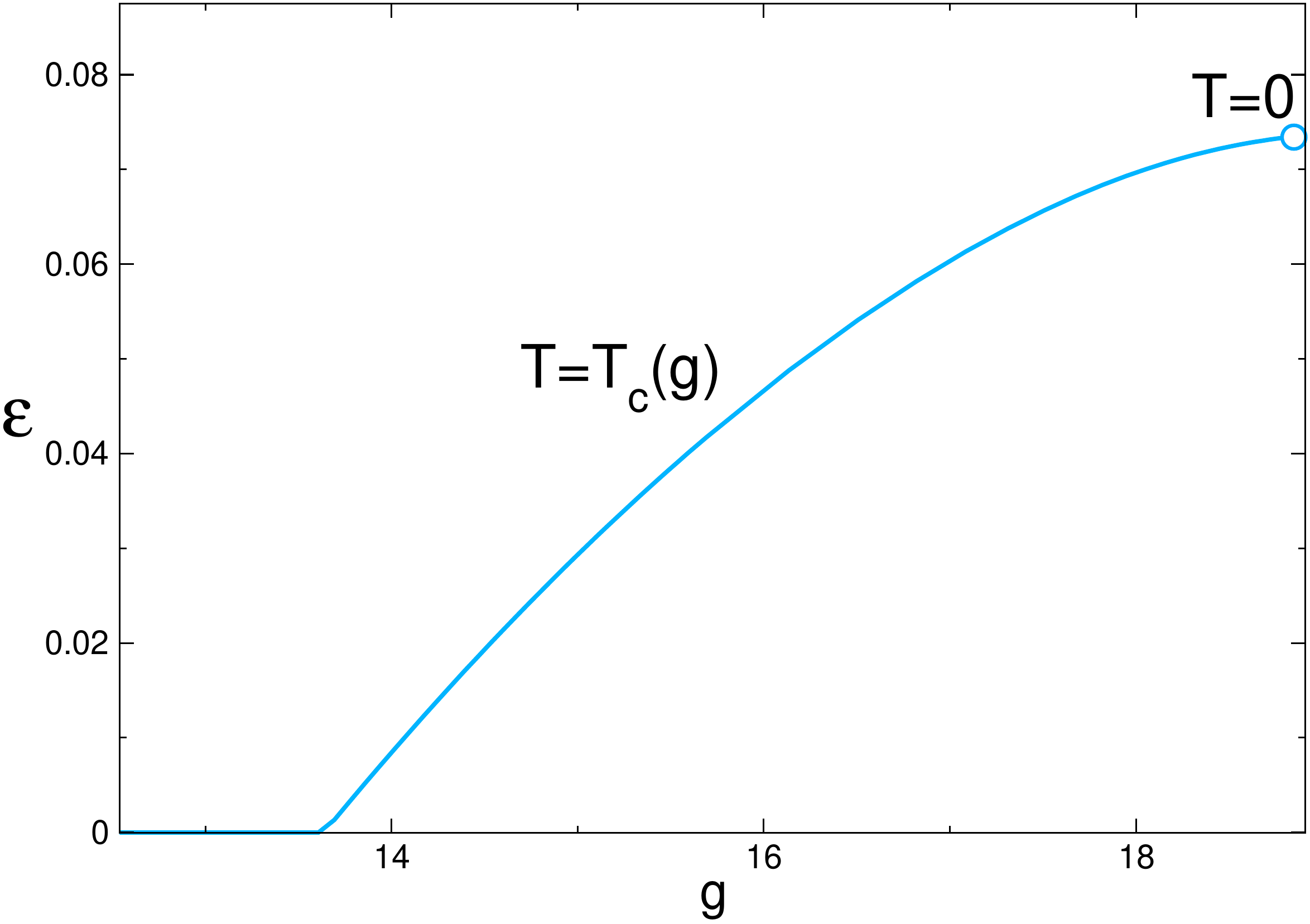}
\caption{ Logarithmic negativity between two nearest-neighbour spins 
 embedded in an infinite system. We show ${\cal E}$ along the critical 
 $T=T_c(g)$ line of the para-ferro transition. Note the sudden death of 
 the negativity at $g\approx 13.6$. 
}
\label{fig:neg_crit}
\end{figure}
%
Several of the generic features of the negativity discussed in 
Fig.~\ref{fig:neg_den} can be extracted by 
considering two adjacent spins embedded in an infinite system. 
In this section we derive analytically the behaviour of the logarithmic 
negativity in this situation. 

\subsubsection{Large $g$ expansion}
\label{sec:two-large-g}

It is straightforward to derive the behaviour of the negativity between the two 
sites in the large $g$ regime (see Fig.~\ref{fig:phadia}). 
The starting point of the analysis are the 
correlators $\langle s_ns_m\rangle$ and $\langle p_np_m\rangle$ in the 
large $g$ limit. These are reported in~\eqref{large-g-1} and~\eqref{large-g-2}. 
Now the matrices $\mathbb{Q}[A]$ and $\mathbb{P}[A^{T_2}]$ (see section~\ref{sec:obs} for 
their definitions) are two-by-two matrices. The negativity spectrum contains only two 
levels. These are the eigenvalues $\nu^2_\pm$ of $\mathbb{Q}[A]\cdot
\mathbb{P}[A^{T_2}]$, which in the large $g$ limit are given as 
\begin{equation}
\label{large-g-eig}
\nu_\pm=\frac{g\pm2}{2g}. 
\end{equation}
Since $\nu_+>1/2$, only $\nu_-$ contributes to the negativity (see~\eqref{neg-boson}), 
which is given as 
\begin{equation}
\label{neg-large-g}
{\cal E}=-\ln\Big(\frac{g-2}{g}\Big). 
\end{equation}
By expanding in the large $g$ limit, one obtains the behaviour ${\cal E}\propto 1/g$. 

\subsubsection{Low-temperature expansion}
\label{sec:two-low-T}

It is also interesting to discuss the limit of low-temperature in the ordered phase 
of the quantum spherical model. To do that, we exploit the expansion of the 
correlators $\langle s_ns_m\rangle$ and $\langle p_np_m\rangle$ 
(cf.~\eqref{low-exp-1} and~\eqref{low-exp-2}). Let us first consider the 
negativity between site $n\equiv(n_x,n_y,n_z)$ and $m\equiv(m_x,m_y,m_z)$. 
The eigenvalues $\nu_i^2$ of $\mathbb{Q}[A]\cdot\mathbb{P}[A^{T_2}]$ are given as 
\begin{align}
	&\nu_1^2=\frac{(W''_n-W''_m)(\sqrt{2}+3(W'_n+W'_m)\beta^2 g)}{12\beta^2g},\\
	&\nu_2^2=\frac{(W'_n-W'_m)(\sqrt{2}\pi^2+15(W''_n+W''_m)\beta^4 g)}{60\beta^4g^2}. 
\end{align}
Here we defined the Watson-type integrals $W_n'$ as 
\begin{equation}
	W_n'\equiv\int\frac{dk}{(2\pi)^3}e^{i k n}\frac{1}{\sqrt{\omega_k}}, \quad
	W_n''\equiv\int\frac{dk}{(2\pi)^3}e^{i k n}\sqrt{\omega_k}.
\end{equation}
Here $\omega_k$ is defined in~\eqref{disp}. Interestingly, the integral $W'_n$ can 
be calculated analytically in terms of hypergeometric functions~\cite{guttmann-2010}. 
We now restrict ourselves to the negativity between two nearest-neighbour spins, i.e., with 
$|n-m|=1$. Specifically,  we choose $n=(0,0,0)$ and $m=(1,0,0,)$. One can 
numerically check that $\nu_1^2<1/4$, implying that $\nu_1$ does not 
contribute to the negativity (see~\eqref{neg-boson}). On the other hand, for a given 
$\beta$, one has that $\nu_2$ contributes to the negativity only for sufficiently large $g$. Specifically, we  
observe that the condition $g>g^*(\beta)$ has to hold, with the ''critical'' value $g^*$ 
given as  
\begin{equation}
\label{gstar}
g^*=\frac{2^{1/4}\pi\sqrt{W'_0-W'_1}}{\sqrt{15}\beta^2\sqrt{1-(W'_0-W'_1)(W''_0+W''_1)}},
\end{equation}
where $W'_0\equiv W'_{(0,0,0)}, W'_1\equiv W'_{(1,0,0)}$, and similarly for $W''$. 
For each value of temperature, the negativity is exactly zero for $g<g^*(T)$. 
Note in~\eqref{gstar} the behaviour $g^*\propto 1/\beta^2$. 
This behaviour persists when considering the negativity between two extended 
systems, although the prefactor in~\eqref{gstar} is different. In particular, 
for two extended systems one obtains  a larger value of $g^*$ in~\eqref{gstar}, 
as it is clear from Fig.~\ref{fig:neg_den}. 

Note that for large enough temperature the negativity vanishes even on the 
critical line (see Fig.~\ref{fig:phadia}). 
This is shown explicitly in Fig.~\ref{fig:neg_crit}. 
The figure shows ${\cal E}$ calculated on the critical 
line $T_c(g)$, with $T_c$ the critical temperature at the given value of $g$. 
The negativity is exactly zero for $g\lesssim 14$, whereas  it is finite 
nonzero up to $g\approx 19$, which corresponds to zero temperature. 

\begin{figure}[t]
\includegraphics[width=0.5\textwidth]{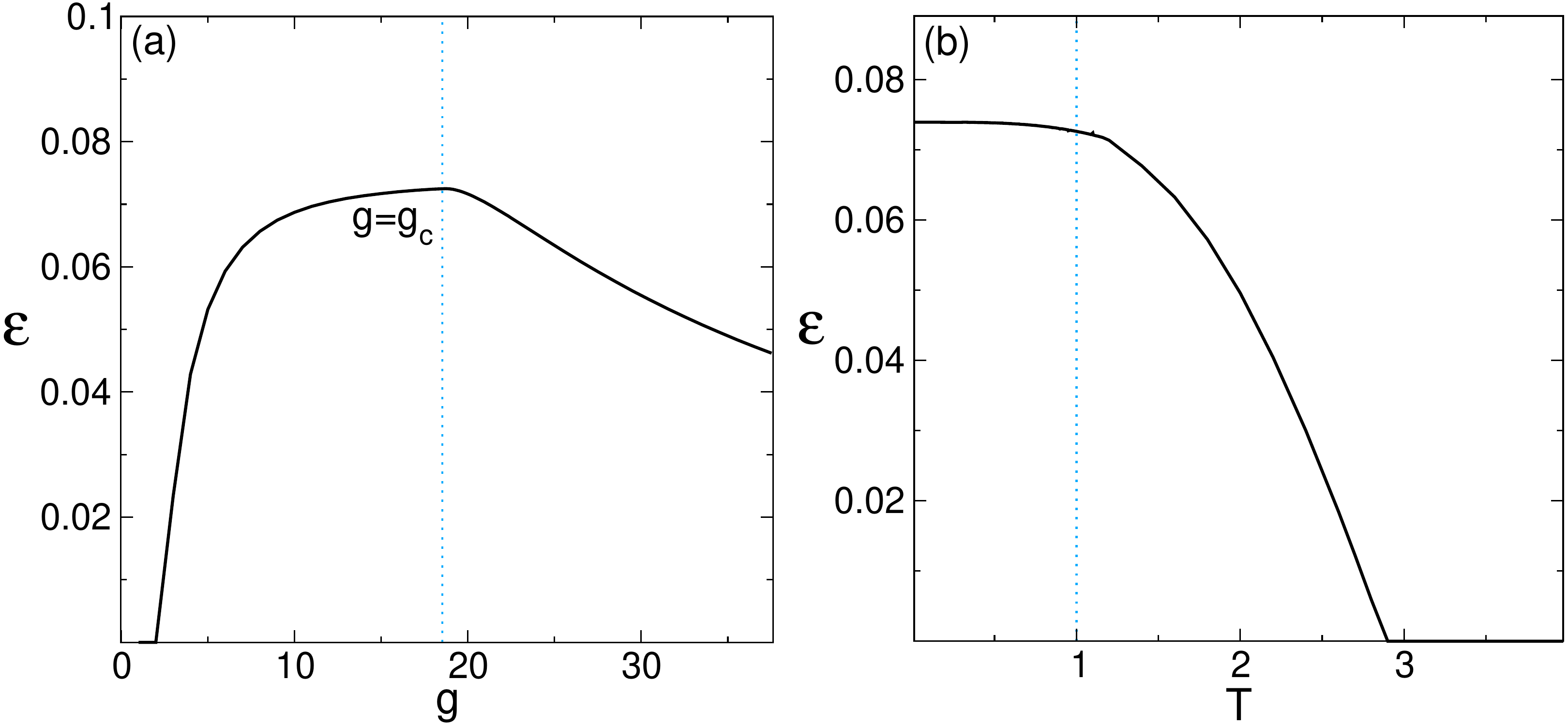}
\caption{ Logarithmic negativity between two adjacent spins in the quantum spherical 
 model: Negativity ${\cal E}$ plotted as a function of $g$ (in (a)) and of the 
 temperature $T$ (in (b)). The figure shows results in the  
 thermodynamic limit. The vertical lines denote the critical point. Data in (a) 
 are for fixed $T=1$. In (b) we fixed $g=g(T_c=1)$ (see 
 Fig.~\ref{fig:phadia}). Note in (a) the slow decay at $g\to\infty$ and the 
 ``death'' of ${\cal E}$ at $g\to0$. Note that in (b) the negativity is 
 identically zero for $T\gtrsim 3$. 
}
\label{fig:neg_two}
\end{figure}

The behaviour of the negativity between two nearest-neighbour sites 
as a function of both temperature and 
quantum coupling $g$ is summarized in Fig.~\ref{fig:neg_two}. 
Panel (a) shows ${\cal E}$ versus $g$, at fixed $T=1$. 
In the figure we plot the negativity and not its density. 
The vanishing behaviour below $g^*\approx 2$, as predicted by Eq.~\eqref{gstar},  
is clearly visible, as  well as 
the slow decay as $1/g$ in the paramagnetic phase. Panel (b) shows ${\cal E}$ 
at fixed $g=g_c(T=1)$ as a function of $T$. The negativity 
exhibits a weak dependence on temperature in the ordered phase, whereas 
it suddenly drops to zero in the paramagnetic phase (sudden death). 

\subsubsection{Critical region}
\label{sec:near-critical}

An important feature in Fig.~\ref{fig:neg_two} (a) is that 
the negativity exhibits regular behaviour across the para-ferro transition. 
On the other hand, recently it has been observed that the negativity can 
exhibit a cusp-like singularity at a finite temperature phase 
transition~\cite{tarun-1,tarun-2}. Moreover, it has been suggested that the 
singular part of the negativity obeys the scaling form as 
\begin{equation}
\label{tarun-sc}
{\cal E}_\mathrm{sing}\propto |g-g_c|^{1-\alpha}, 
\end{equation}
with $\alpha$ the specific heat critical exponent. 
Our results are consistent with Ref.~\onlinecite{tarun-1,tarun-2}. 
Specifically, for the $3D$ QSM one has $\alpha=-1$. 
Thus, Eq.~\eqref{tarun-sc} predicts that ${\cal E}_\mathrm{sing}$ has a 
singular term $(g-g_c)^2$, i.e., a weaker than a cusp singularity. 
On the other hand, for $\alpha=0$, which is the case investigated in 
Ref.~\onlinecite{tarun-2}, one has the cusp-like behaviour as 
${\cal E}_\mathrm{sing}\propto |g-g_c|$. 

To clarify the behaviour of ${\cal E}$ for two nearest-neighbour sites, 
it is useful to consider the single-particle negativity spectrum. 
Using the expansions for the correlators near the critical point 
(cf.~\eqref{snsm-crit-g}\eqref{snsm-crit-beta}\eqref{pnpm-crit-beta}, 
and~\eqref{pnpm-crit-g}), it is straightforward to obtain the 
spectrum levels. These are given as 
\begin{align}
&\nu_1^2=(S_{nn}^{(0)}-S_{nm}^{(0)})(P_{nn}^{(0)}+P_{nm}^{(0)})+
\big[(S_{nn}^{(1)}-S_{nm}^{(1)})(P_{nn}^{(0)}
		+P_{nm}^{(0)})+
		(S_{nn}^{(0)}-S_{nm}^{(0)})(P_{nn}^{(1)}
		-P_{nm}^{(1)})
\big](g-g_c),\\
&\nu_2^2=(S_{nn}^{(0)}+S_{nm}^{(0)})(P_{nn}^{(0)}-P_{nm}^{(0)})+
\big[(S_{nn}^{(1)}+S_{nm}^{(1)}+2Z_\mathrm{sing})(P_{nn}^{(0)}
	-P_{nm}^{(0)})+(S_{nn}^{(0)}+S_{nm}^{(0)})(P_{nn}^{(1)}
	+P_{nm}^{(1)})
\big](g-g_c). 
\end{align}
Here we defined 
\begin{align}
\label{lev-1}
	S_{nm}^{(0)} & \equiv 
\frac{\sqrt{g_c}}{2\sqrt{2}}
	\int\frac{dk}{(2\pi)^3}\frac{\coth(\beta\sqrt{g_c\omega_k/2})}{\sqrt{\omega_k}}
	e^{ik(n-m)}
\\
\label{lev-2}
S_{nm}^{(1)} & \equiv 
\int \frac{dk}{(2\pi)^3}\Big(\frac{\beta}{8}+
	\frac{\mathrm{coth}(\beta\sqrt{g_c\omega_k/2})}{4\sqrt{2g_c\omega_k}}-\frac{\beta}{8}
	\mathrm{coth}^2(\beta\sqrt{g_c\omega_k/2})\Big)e^{ik(n-m)}\\
	\label{lev-3}
	P_{nm}^{(0)} & \equiv
	\frac{1}{\sqrt{2g_c}}
	\int\frac{dk}{(2\pi)^3}\coth(\beta\sqrt{g_c\omega_k/2})\sqrt{\omega_k}
	e^{ik(n-m)}\\
\label{lev-4}
	P_{nm}^{(1)} &\equiv 
\frac{1}{g_c}\int\frac{dk}{(2\pi)^3}
	\omega_k
\Big(\frac{\beta}{4}-
	\frac{\mathrm{coth}(\beta\sqrt{g_c\omega_k/2}}{2\sqrt{2g_c\omega_k}}-\frac{\beta}{4}
	\mathrm{coth}^2(\beta\sqrt{g_c\omega_k/2})\Big)e^{ik(n-m)}\\
\label{lev-5}
Z_\mathrm{sing} & \equiv
2\sqrt{2}\pi\beta\int \frac{dk}{(2\pi)^3}\Big(\frac{\beta}{8}+\frac{\mathrm{coth}(\beta\sqrt{g_c\omega_k/2})}{4\sqrt{2g_c\omega_k}}-\frac{\beta}{8}
	\mathrm{coth}^2(\beta\sqrt{g_c\omega_k/2})\Big)
\end{align}
%
\begin{figure}[t]
\includegraphics[width=0.5\textwidth]{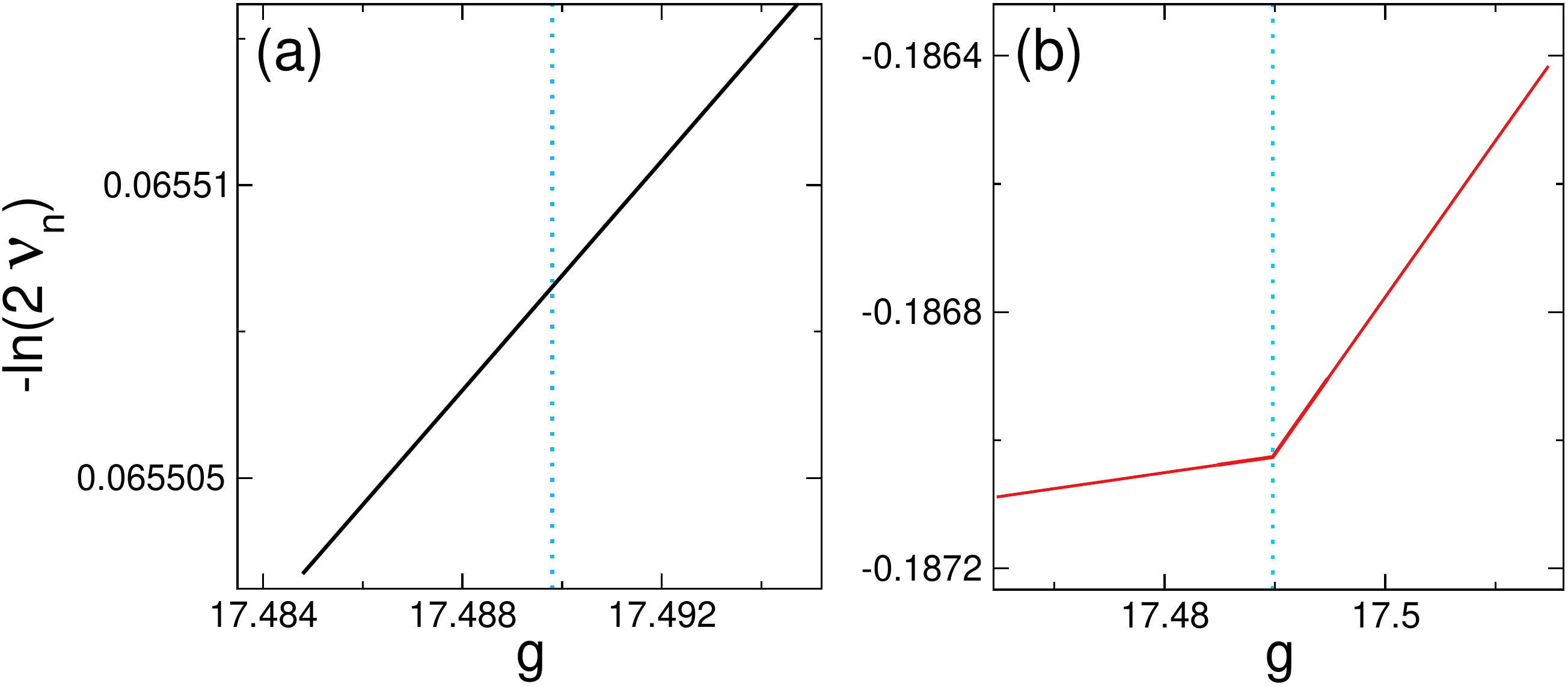}
\caption{ Single-particle negativity spectrum for two adjacent spins the 
 embedded in an infinite system. We plot $-\ln(2\nu_n)$, with $\nu_n^2$ the eigenvalues 
 of the correlation matrix, versus $g$. Data are for fixed  $T\approx 1.486$. 
 The spectrum consists of two eigenvalues 
 shown in (a) and (b). Note that in (b) $-\ln(2\nu_n)<0$, implying that  the level 
 does not contribute to the negativity. Note the kink at the critical point (vertical 
 line). In contrast, the eigenvalue in (a) is regular, implying that the negativity 
 is not singular at the critical point. 
}
\label{fig:two_eig}
\end{figure}
%
Note that $Z_\mathrm{sing}$ contains the singular term $\sqrt{\mu}$ that 
we derived in section~\ref{sec:mu}. Interestingly, this 
affects only $\nu_2$, whereas $\nu_1$ is regular. 

In Fig.~\ref{fig:two_eig} we show the levels of the single-particle 
negativity spectrum (cf.~\eqref{lev-1} and~\eqref{lev-2}). In the 
figure we plot $-\ln(2\nu_n)$ versus $g$. In (a) we show the regular 
eigenvalue $\nu_1$, whereas $\nu_2$ is reported in (b). The singularity 
as $|g-g_c|$ in (b) is clearly visible. Importantly, one has that $-\ln(2\nu_2)<0$, 
implying that $\nu_2$ does not contribute to the logarithmic negativity. 
This suggests that in the case of two extended subsystems 
only a subset of the single-particle negativity spectrum 
levels will exhibit singular behaviour. 
These levels, however, do not contribute to the negativity, which is regular 
at the finite-temperature transition. This point will be better 
clarified when we will discuss the negativity spectrum of two extended 
regions in section~\ref{sec:neg-spect}.

\subsection{Half-system negativity}
\label{sec:half}

\begin{figure}[t]
\includegraphics[width=0.9\textwidth]{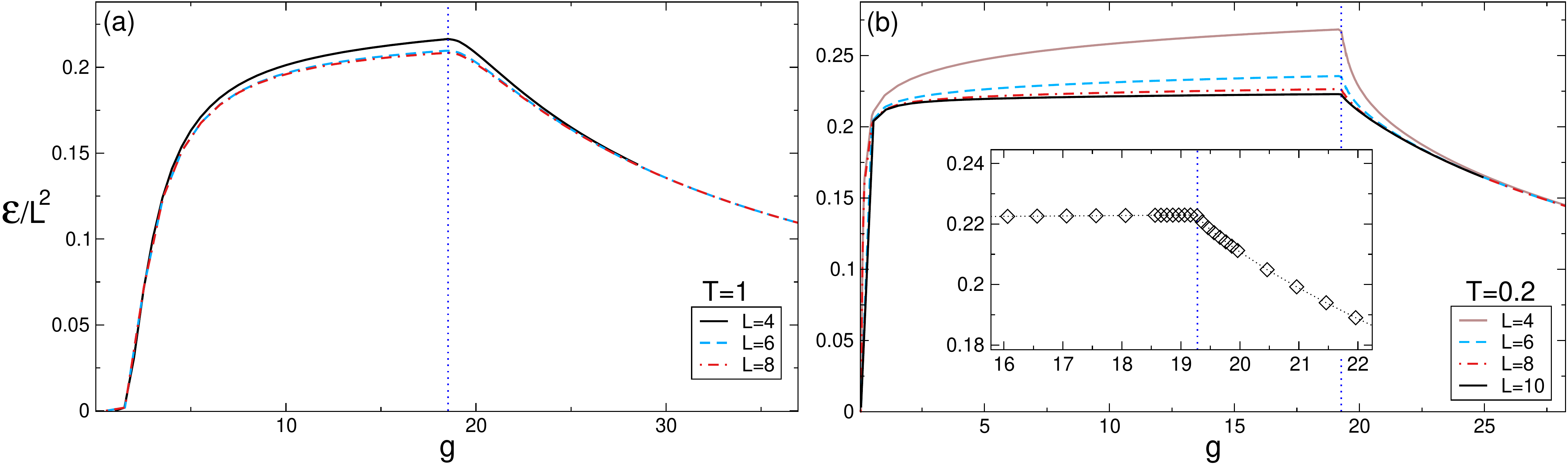}
\caption{ Logarithmic negativity in the finite-temperature quantum spherical 
 model. (a) and (b) show the negativity density ${\cal E}/L^2$ plotted versus 
 the quantum coupling $g$ for $T=1$ and $T=0.2$, respectively. Data are obtained 
 by using the thermodynamic result for the spherical constraint $\mu$. 
 The effect of the system size $L$ is small, especially for $T=1$. The qualitative 
 behaviour is similar to Fig.~\ref{fig:neg_two}  
 (compare (a) and (b)). For $T=1$ the negativity exhibits regular behaviour 
 at $g_c$, whereas in the limit $T\to0$ a cusp-like feature appears at  
 $g_c$. In (b) the inset shows the data for $L=10$, zooming around the critical point. 
}
\label{fig:neg2}
\end{figure}

We now discuss the half-system logarithmic negativity. 
Our results are reported in Fig.~\ref{fig:neg2}. The figure provides an overview 
of ${\cal E}$ at fixed temperature $T=1$ and $T=0.2$ ((a) and (b), respectively) 
as a function of $g$. In both panels we plot the negativity density ${\cal E}/L^2$ 
of half system. The different lines are for different sizes $L$. Similar to 
section~\ref{sec:free-e}, we calculate the negativity by using the finite-size 
result for the correlators (see~\eqref{snsm} and~\eqref{pnpm}), 
and the value of the spherical parameter $\mu$ in the thermodynamic limit.  
To regularize the divergent contribution of the zero mode we fix $\mu=10^{-6}$ 
for $g\le g_c$. We also checked that the results for the negativity do not 
depend on the choice of the regularization.  

Clearly, in Fig.~\ref{fig:neg2} finite-size effects are small, especially for 
$T=1$ (see (a)). In (a) the data for $L=8$ cannot be distinguished 
from the result in the thermodynamic limit. In both (a) and (b), the negativity 
has its maximum value at the critical point. In the paramagnetic phase the negativity 
decays as $1/g$, similar to the case of two spins (see section~\ref{sec:two-spin}). 
Within the ordered phase ${\cal E}$ exhibits a mild dependence on $g$, except at 
small $g$, where it drops dramatically. Specifically, for 
$g<g^*\approx 2$ the negativity is exactly zero. Note that  $g^*$ decreases with decreasing the 
temperature (compare (a) and (b) in the figure). This behaviour is similar to 
what observed for the two-site 
negativity (see Fig.~\ref{fig:neg_two}). Again, this implies that for any value of 
the temperature there is a ``critical'' value of the coupling $g$ below which 
the quantum fluctuations are not strong enough to give a finite 
entanglement negativity. An important remark, however, is that since 
the negativity gives only a bound on the entanglement, the vanishing of ${\cal E}$ 
below $g^*$ does not imply the absence of entanglement.  
From Fig.~\ref{fig:neg2} (a) it is clear that at $T=1$, 
the negativity exhibits regular behaviour  at  $g_c$. 
On the other hand, upon decreasing the temperature the negativity 
develops a cusp. This is clear from Fig.~\ref{fig:neg2} (b). 
Similar (weak) singular behaviour was observed in the 
area-law prefactor of the von Neumann entropy in ground-state quantum 
phase transitions~\cite{kallin-2013,helmes-2014,frerot-2016}.

\begin{figure}[t]
\includegraphics[width=0.5\textwidth]{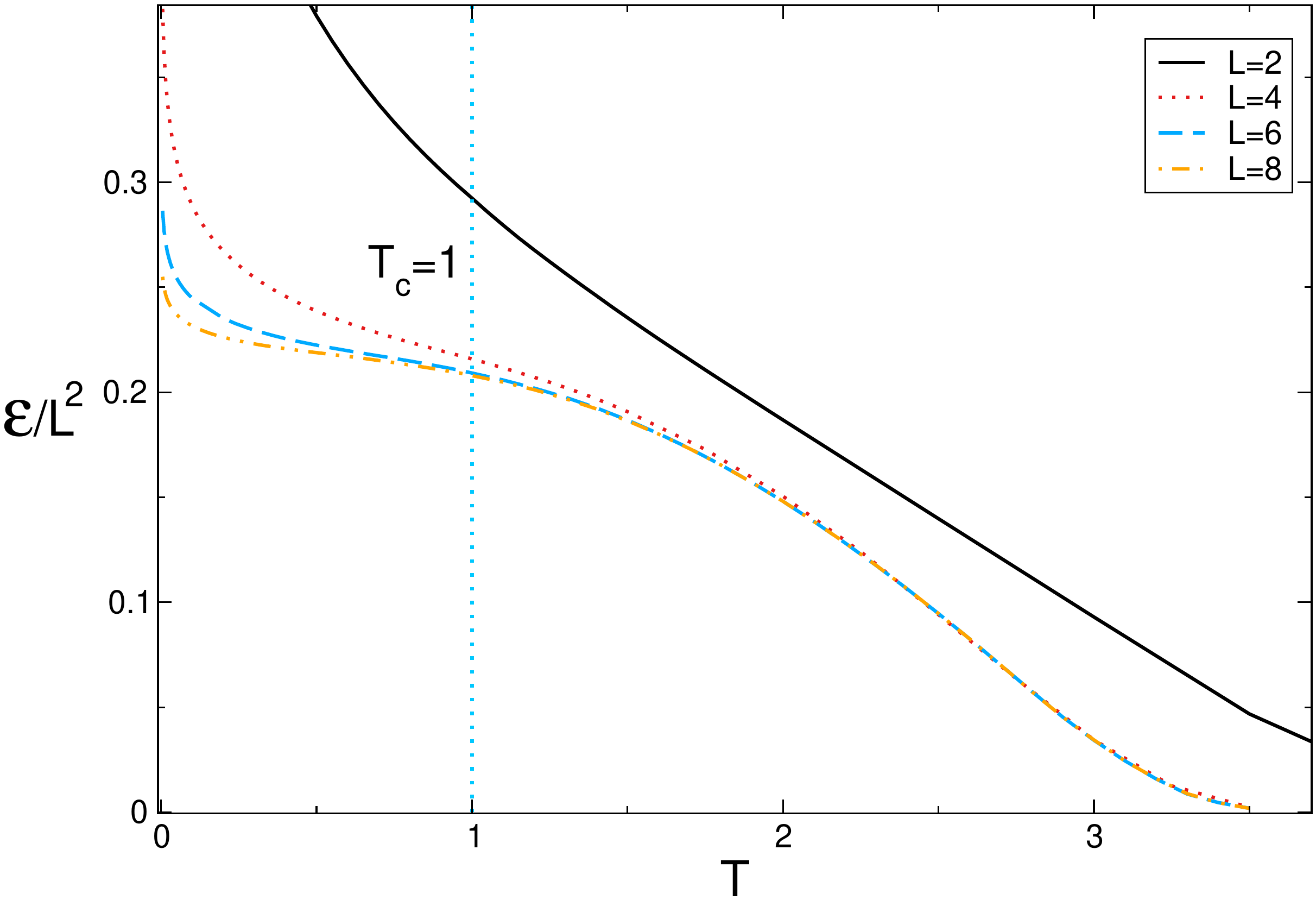}
\caption{ Density of the logarithmic negativity ${\cal E}/L^2$ in the three-dimensional 
 quantum spherical model at fixed $g_c\approx18.52$ plotted as a function of temperature $T$. 
 Data are obtained by using the value of the spherical parameter $\mu$ in the 
 thermodynamic limit. 
 The vertical line is the critical temperature $T_c=1$ at  $g_c$ (see 
 Fig.~\ref{fig:phadia}). Note that ${\cal E}=0$ for $T\gtrsim\approx3.5$. 
 In the limit $T\to0$ quantum fluctuations are enhanced and ${\cal 
 E}/L^2$ increases, although area-law ${\cal E}\propto L^2$ behaviour 
 should persist even at zero temperature. 
}
\label{fig:neg_T}
\end{figure}

It is interesting to investigate the behaviour of the logarithmic 
negativity at fixed $g$, i.e., approaching criticality by changing the temperature.  
This is illustrated in Fig.~\ref{fig:neg_T}. We plot ${\cal E}/L^2$ 
versus $T$ for fixed $g=g_c(T=1)$. 
Now the negativity exhibits a fast decay with $T$ in the paramagnetic 
region, and already for $T\approx 3$ it is 
exactly zero. This is in contrast with the result at fixed $T$ (see 
Fig.~\ref{fig:neg2}), where the behaviour as $1/g$ is observed. 
Finite-size effects are larger upon decreasing the temperature 
in the ordered phase. Note that for the smaller system sizes one has a quite 
large value of ${\cal E}$ in the limit $T\to0$, although the area-law 
behaviour ${\cal E}\propto L^2$ is expected to hold also at zero temperature. 
Finally, the negativity is not singular at the critical 
point, as in Fig.~\ref{fig:neg2}. 

\begin{figure}[t]
\includegraphics[width=0.5\textwidth]{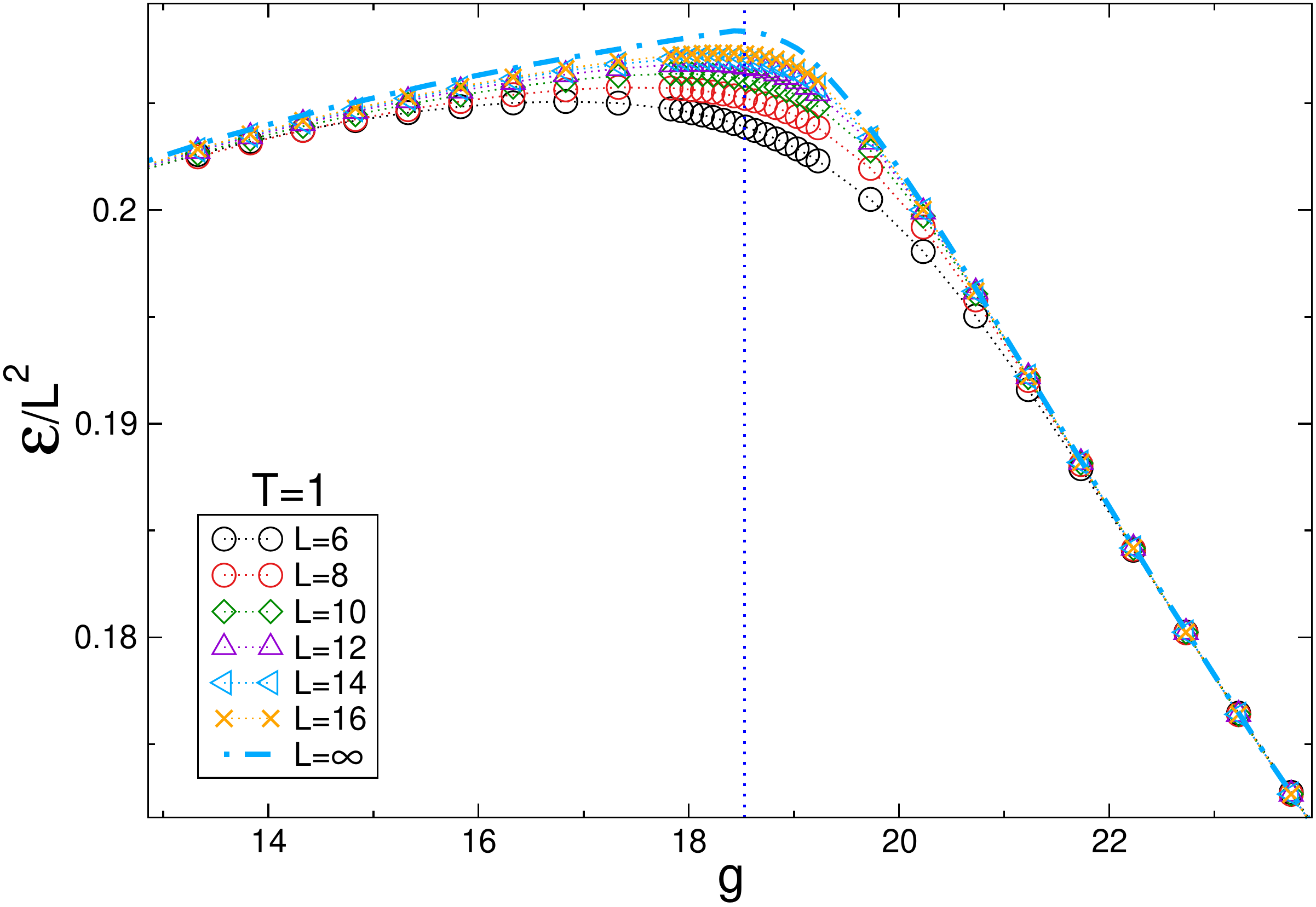}
\caption{ Scaling of the negativity ${\cal E}$ at a finite-temperature 
 quantum phase transition in the quantum spherical model. We plot ${\cal E}/L^2$ 
 versus $g$. The results are at fixed $T=1$. Different symbols are for different 
 sizes $L$. Data are obtained by solving the finite-size constraint equation~\eqref{mu-fs}. 
 The vertical dotted line marks the critical point at $g_c\approx 18.52$. 
 The dashed-dotted line is the result in the thermodynamic limit. Note the larger 
 finite-size effects in the critical region. 
}
\label{fig:negativityL}
\end{figure}

Finally, we investigate the behaviour of the logarithmic negativity in 
finite-size systems. To address this point,  we provide results for the 
half-system negativity obtained by using the finite-size value of the spherical 
parameter $\mu$, i.e., by numerically solving~\eqref{mu-fs}. 
Our results are reported in Fig.~\ref{fig:negativityL}. The figure plots 
the negativity density ${\cal E}/L^2$ versus $g$, for several sizes (symbols 
in the figure). Similar to the mutual informations (cf. Fig.~\ref{fig:mi} 
and~\ref{fig:mi-n2}), the data show small finite-size effects both in the paramagnetic 
phase and in the ordered phase. At the critical point (vertical line) finite-size 
effects are larger. However, the data appear to converge to the 
thermodynamic limit result
(dashed-dotted line). The latter is obtained from Fig.~\ref{fig:neg2}. 
Note that the data for different $L$s do not exhibit any crossing  at the 
critical point. 

\section{Entanglement spectra, negativity spectra, and the zero mode}
\label{sec:neg-spect}

\begin{figure}[t]
\includegraphics[width=0.5\textwidth]{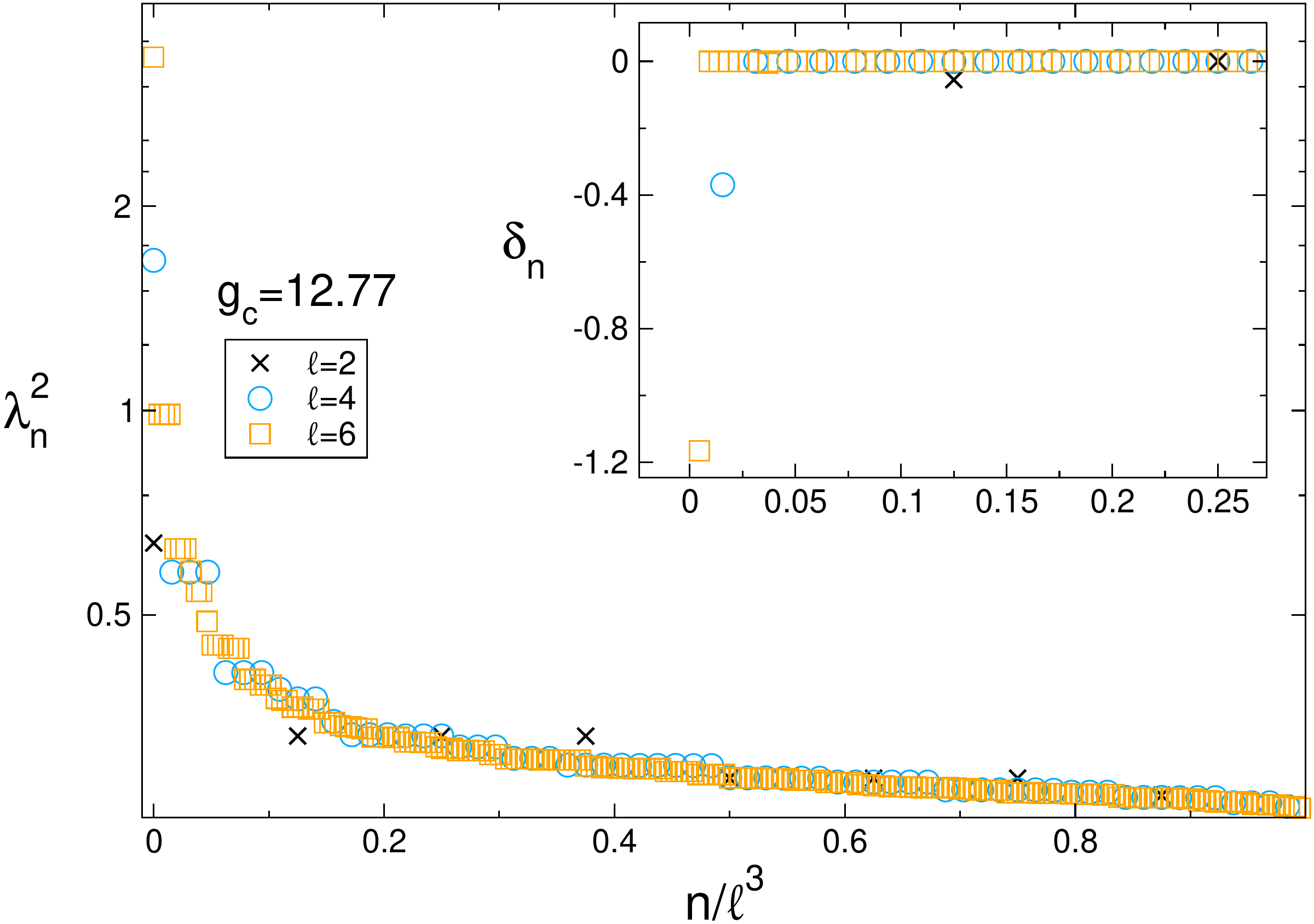}
\caption{ Single-particle  entanglement spectrum between two adjacent blocks 
 embedded in an infinite system. The figure shows the single-particle 
 entanglement spectrum levels $\lambda_n^2$. The results are at the critical point at $g_c\approx 
 12.77$. Subsystem $A$ consists of a cube of size $\ell$. Different symbols are for different 
 $\ell$. A logarithmic scale is used on the $y$-axis. Inset: $\delta_n\equiv (\lambda_n^2)'_+-
 (\lambda_n^2)'_-$, with $(\lambda_n^2)'_\pm$ the left and right derivative of 
 $\lambda_n^2$ with respect to $g$ calculated at $g_c$. Data are for 
 $\ell=6$. Note that $\delta_n\ne 0$ only for a small subset of levels. 
}
\label{fig:ent_cube}
\end{figure}

\begin{figure}[t]
\includegraphics[width=0.5\textwidth]{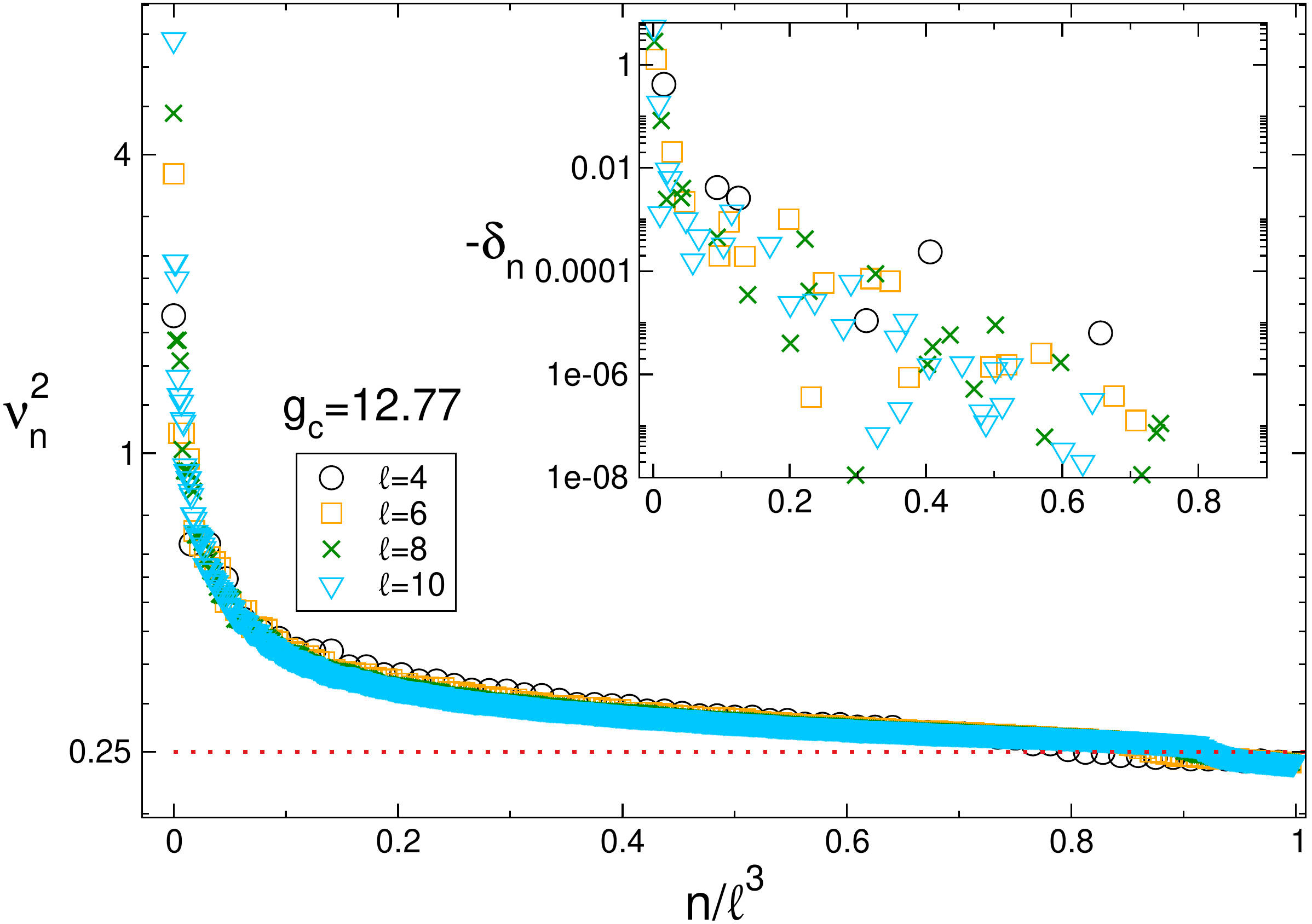}
\caption{ Single-particle  negativity spectrum between two adjacent blocks embedded in an 
 infinite system. The figure shows the single-particle negativity spectrum levels 
 $\nu_n^2$. Subsystem $A$ is a cube of size $\ell$. $A_1$ and $A_2$ are the two halves of the 
 cube of linear sizes $\ell_x=\ell/2,\ell_y=\ell,\ell_z=\ell$ (see Fig.~\ref{fig:cartoon} (c)). 
 The results are at the critical point at $g_c\approx 
 12.77$. Note that only the levels $\nu_n^2<1/4$ (horizontal dashed line) 
 contribute to the logarithmic negativity. Inset: $\delta_n\equiv (\nu_n^2)'_+-
 (\nu_n^2)'_-$, with $(\nu_n^2)'_\pm$ the left and right derivative of 
 $\nu_n^2$ with respect to $g$ at the critical point $g_c$. 
 Note that $\delta_n=0$ for large $n$, i.e., for the levels that 
 contribute to the negativity. 
}
\label{fig:neg_cube}
\end{figure}

In section~\ref{sec:near-critical} we showed that the negativity 
spectrum for two nearest-neighbor sites contains one regular and one singular 
levels. The latter exhibits a cusp-like singularity across the finite temperature 
transition. The singular 
level, however, does not contribute to the logarithmic negativity. Moreover, 
we observed in section~\ref{sec:half} that the negativity between two 
extended blocks shows regular behaviour across the transition, similar to 
the entanglement entropy (see Fig.~\ref{fig:entropy}). It is interesting 
to investigate how this is reflected in the negativity spectrum of two 
extended regions. It is also interesting to compare the singularity structure 
of the negativity spectrum and the entanglement spectrum.

\subsection{Entanglement spectra}
\label{sec:e-spectrum}

Let us start discussing the entanglement spectrum. 
Let us consider the two matrices $\mathbb{Q}_{nm}=
\langle s_n s_m\rangle$ and 
$\mathbb{P}_{nm}=\langle p_np_m\rangle$ (see section~\ref{sec:obs}). 
Here we work in the thermodynamic limit. 
Since we are interested in the cusp-like singularity of the 
entanglement spectrum, it is convenient to expand 
$\mathbb{Q}$ and $\mathbb{P}$ as 
\begin{align}
\label{expa}
\mathbb{Q}=\mathbb{Q}_0+(g-g_c)\mathbb{Q}_1^{\pm}+\dots\\
\label{expa-1}
\mathbb{P}=\mathbb{P}_0
+(g-g_c)\mathbb{P}_1+\dots,
\end{align}
where the dots denote higher order terms in powers of $g-g_c$, which we neglect. 
The matrices $\mathbb{Q}_0$ and $\mathbb{P}_0$ are obtained from $S_{nm}^{(0)}$ 
and $P_{nm}^{(0)}$ (cf.~\eqref{lev-1} and~\eqref{lev-3}), whereas  $\mathbb{Q}_1$ and 
$\mathbb{P}_1$ are easily derived from $S_{nm}^{(1)}$ 
and $P_{nm}^{(1)}$ (see~\eqref{lev-2} and~\eqref{lev-4}). 
The $\pm$ in $\mathbb{Q}_1^\pm$ is to stress that the matrix is different 
on the two sides of the transition, because of the term $Z_\mathrm{sing}$ 
(see~\eqref{lev-5}), which is only present for $g>g_c$. We now have at the 
leading order ${\mathcal O}(g-g_c)$ 
\begin{equation}
\label{q-first}
\mathbb{Q}\cdot\mathbb{P}=\mathbb{Q}_0\cdot
\mathbb{P}_0+(g-g_c)(\mathbb{Q}_0\cdot
\mathbb{P}_1+\mathbb{Q}_1^\pm\cdot
\mathbb{P}_0)+\dots
\end{equation}
By using standard perturbation theory, one obtains that the 
corrections to the eigenvalues $\lambda_n^2$ of the matrix $\mathbb{Q}_0\cdot
\mathbb{P}_0$ are given as 
\begin{equation}
\label{delta-nu}
\delta\lambda_n^2=(g-g_c)\langle\phi_n|\mathbb{Q}_0\cdot
\mathbb{P}_1+\mathbb{Q}_1^\pm\cdot
\mathbb{P}_0|\phi_n\rangle,
\end{equation}
where $|\phi_n\rangle$ is the eigenvector of $\mathbb{Q}_0\cdot\mathbb{P}_0$ 
corresponding to eigenvalue $\lambda_n^2$. 
Clearly, the singularity in the negativity is determined by the second term 
in~\eqref{delta-nu}, which depends on $Z_\mathrm{sing}$. 
Now we observe that $Z_\mathrm{sing}$ does not depend on the positions 
$n,m$. This has striking consequences for the single-particle entanglement 
spectrum. First, the singular contribution in~\eqref{delta-nu} is given as 
\begin{equation}
\label{delta-sing}
[\delta\lambda_n^2]_\mathrm{sing}=(g-g_c)
Z_\mathrm{sing}\langle\phi_n|(1,1,\dots)\otimes(1,1,\dots)\cdot\mathbb{P}|\phi_n\rangle. 
\end{equation}
Note that the vector $(1,1,\dots)\otimes(1,1,\dots)\cdot \mathbb{P}|\phi_n\rangle$ 
is flat, i.e, all the elements are equal. Moreover, we numerically observed that the 
eigenvector $|\phi_0\rangle$ is approximately flat, i.e., all its components are $1/\ell^{3/2}$, 
which reflects the presence of a zero mode.  
This implies that for $n=0$ the expectation value in~\eqref{delta-sing} is nonzero. 
On the other hand, 
the components of $|\phi_n\rangle$ with $n>0$ are real and are orthogonal 
to $|\phi_0\rangle$. 
This implies that the expectation value 
in~\eqref{delta-sing} approximately vanishes for $n>0$. This allows us to conclude  
that the only few spectrum levels  are singular across the transition, i.e., 
the ones related  to the zero mode. 
We summarize our results in Fig.~\ref{fig:ent_cube}. The figure shows the single-particle 
entanglement spectrum levels $\lambda_n^2$. Here subsystem $A$ is the cube  of 
size $\ell$ (see Fig.~\ref{fig:cartoon} (c)) embedded in an infinite system. 
One has that  $\lambda^2_n>1/4$ $\forall n$. The eigenvalues $\lambda_n^2$ quickly 
decay with their index $n$. For most 
of the eigenvalues, one has  $\lambda_n^2\approx 1/4$. In the inset of 
Fig.~\ref{fig:ent_cube} we plot $\delta_n$ defined as 
\begin{equation}
\label{dn-def}
\delta_n\equiv(\lambda_n^2)'_+-(\lambda_n^2)'_-.
\end{equation}
Here $(\lambda_n^2)'_\pm$ are the right and left derivatives with respect to $g$ of 
$\lambda_n^2$, calculated at $g_c$. 
A nonzero value of $\delta_n$ signals that the level $\lambda_n$ is singular 
across the transition. 
The results in the figure are obtained by 
using~\eqref{delta-sing}. Clearly, one has that $\delta_n\ne0$ only for small $n$.

\subsection{Negativity spectra}
\label{sec:n-spectrum}

We now discuss the negativity spectrum.  Subsystem $A$ is now divided into two 
parts $A_1,A_2$ of sizes $\ell_x=\ell/2,\ell_y=\ell,\ell_z=\ell$ 
(see Fig.~\ref{fig:cartoon} (c)). The partial 
transposition is performed with respect to $A_2$. 
Our results for the negativity spectrum are shown in Fig.~\ref{fig:neg_cube}. 
In the main figure we show 
the eigenvalues $\nu_n^2$ of the matrix $\mathbb{Q}_0\cdot\mathbb{P}_0[A^{T_2}]$ 
versus $n/\ell^3$. For most of the eigenvalues one has 
that $\nu_n^2>1/4$, implying that they do not contribute to the negativity 
(see section~\ref{sec:ent-cal}). As for the entanglement spectrum, 
the flat vector is an approximate eigenvector of $\mathbb{Q}_0\cdot
\mathbb{P}_0[A^{T_2}]$, with eigenvalue $\nu_0^2$. 
For instance, we numerically checked that for $\ell=4,6$ the eigenvector associated 
with $\nu_0^2$ has components are in the interval $[-0.15,-0.10]$, clustering  around 
the expected value $1/\ell^{3/2}=0.125$. This behaviour is reflected in that 
of $\delta_n$. The definition of $\delta_n$ is the same as in~\eqref{dn-def} after 
replacing $\lambda_n\to\nu_n$. Clearly, the expansions~\eqref{expa} also hold 
upon replacing $\mathbb{P}_0\to\mathbb{P}_0[A^{T_2}]$ and $\mathbb{P}_1\to 
\mathbb{P}_1[A^{T_2}]$. For each negativity spectrum level, $\delta_n$ is plotted 
in the inset in Fig.~\ref{fig:neg_cube}. In contrast with the entanglement spectrum 
(compare with Fig.~\ref{fig:ent_cube}), one has $\delta_n\ne0$ for a large subset of 
levels at small $n$. For instance, 
for $\ell=6$ one has $\delta_n=0$ only for $n\gtrsim 10$. As it is clear from 
Fig.~\ref{fig:ent_cube} all the levels of the negativity spectrum with $\delta_n\ne 0$ 
do not contribute to the negativity because they correspond to $\lambda_n^2>1/4$. 
This suggests a suppression of the logarithmic correction to the boundary-law scaling 
of the logarithmic negativity, in contrast with the entanglement entropy (see 
section~\ref{sec:e-spectrum}).

\section{Conclusions}
\label{sec:concl}

We investigated the interplay between entanglement and classical fluctuations 
at finite-temperature critical points. Specifically, we focused on the three 
dimensional quantum spherical model, which has a finite-temperature transition 
between a paramagnetic 
phase and a ferromagnetically ordered phase. We considered several entanglement-related 
observables, such as the von Neumann and R\'enyi entropies, the mutual information, 
and the logarithmic negativity. In particular, we characterized the behaviour of the 
logarithmic negativity in all the different phases and at the transition. 
We also investigated how the behaviour of the entropies and of the negativity is 
reflected in the single-particle entanglement spectrum and on the negativity 
spectrum. 

We now mention several important directions for future research. First, it would be 
important to explore the behaviour of entanglement-related observables in different 
dimensions. For instance, in $d=4$ the para-ferro transition becomes mean field. It 
would be interesting to investigate how this is reflected in the singularity structure 
of entanglement. An interesting direction is to investigate the 
logarithmic negativity at the quantum phase transition in $d=3$. Our results 
suggest that the logarithmic negativity exhibits a cusp-like singularity. It would 
be useful to understand how this is reflected in the structure of the negativity 
spectrum. Another interesting direction is to study the crossover from 
quantum to classical criticality. 
Moreover, it would be enlightening to generalize the perturbative analysis developed 
in section~\ref{sec:neg-spect} including higher-order corrections. This would allow, 
in principle, to characterise the full singularity structure of entanglement-related 
quantities. 

It would be also useful to extend our analysis to other entanglement-related 
quantities, such as the quantum Fisher information, especially in the light of the results of 
Ref.~\onlinecite{hauke-2016}. 
Interestingly, the quantum spherical model remains exactly solvable even in the presence 
of long-range interactions. This opens the possibility of studying entanglement-related 
quantities in long-range models. 

Another important direction is to study 
the out-of-equilibrium dynamics of entanglement-related quantities after a quantum 
quench. In recent years, it has been shown~\cite{CCsemiclassics,FaCa08,AlCa17,AlCa17a} that for integrable models it is possible 
to describe quantitatively the entanglement dynamics by combining integrability with 
a quasiparticle picture. It would be interesting to investigate the validity of 
this picture for $d>1$. In $d=3$ this could also allow to investigate whether the 
presence of finite-temperature criticality affects the entanglement dynamics. Finally, 
it has been suggested in Ref.~\cite{gge-mf} that in the $O(N)$ model with $N\to\infty$ 
the steady-state arising after a quantum quench is not described by the so-called 
Generalized Gibbs Ensemble. The quantum spherical model provides an ideal framework 
to clarify this issue. 

\section{Acknowledgements}
\label{sec:ack}

We thank F. Parisen Toldin for several useful discussions. 
VA acknowledges  support  from  the  D-ITP  consortium,  a  program  
of  the  NWO, and from the European Research Council under the 
ERC Advanced grant 743032 DYNAMINT.

\appendix

\section{Evaluation of some Watson-type integrals}
\label{sec:app}

The numerical evaluation of the three-dimensional integrals appearing in the 
expression for the correlators in~\eqref{snsm} is a demanding task. 
However, there are well-known tricks to reduce 
them to one-dimensional integrals. To proceed we imagine of expanding 
the integrand in~\eqref{snsm} in the 
limit $k\to0$. Thus, one obtains that the first term is $\propto1/\omega_k$. 
The higher-order terms yield integrals of the type 
$\int dk/(2\pi)^3\omega_k^\alpha$, with $\alpha$ a positive integer. 
These integrals can be performed analytically by exploiting the fact that~\cite{series-2000}  
\begin{multline}
\label{nice-green}
\int_{-\pi}^{\pi}\frac{dk}{(2\pi)^3}e^{ik x}\big(-\sum_{j=x,y,z}\cos k_j\big)^\alpha=
\\\left\{
\begin{array}{cc}
0 & \textrm{if}\, \alpha=0,\dots,X-1\\
(-1)^\alpha\alpha! 2^{-|x_1|-|x_2|-|x_3|}\sum_{p_1,p_2,p_3\ge 0} 2^{-2p_1-2p_2-2p_3}
\frac{\delta_{p_1+p_2+p_3,(\alpha-X)/2}}{p_1!p_2!p_3!(p_1+|x_1|)!(p_2+|x_2|)!(p_3+|x_3|)!}
&\textrm{if}\, \alpha-X\ge 0,\alpha-X \textrm{even}.
\end{array}
\right.
\end{multline}
Here we defined $X\equiv |x_1|+|x_2|+|x_3|$. 
Clearly, the integral of $\omega_k^\alpha$ can be obtained by using 
Eq.~\eqref{nice-green} and Newton's binomial formula. 
We now discuss the integration of the singular contribution. The resulting 
integral is of the type 
\begin{equation}
\label{s00-int}
\int\frac{dk}{(2\pi)^3}e^{ik(n-m)}\omega_k^{-1}.
\end{equation}
%
The integral~\eqref{s00-int} 
can be rewritten as a  one-dimensional integral. One can use  the identity
\begin{equation}
\label{id-1}
\frac{1}{1+z}=\int_0^\infty dt e^{-(z+1)t}. 
\end{equation}
After using~\eqref{id-1} in~\eqref{s00-int}, the integration over $k$ can be 
performed explicitly. One obtains 
\begin{equation}
\label{id-2}
	\int\frac{dk}{(2\pi)^3}\frac{e^{ik(n-m)}}{\omega_k}=
	\int_0^\infty dt e^{-3t}I_{n_x-m_x}(t)I_{n_y-m_y}(t)I_{n_z-m_z}(t),
\end{equation}
where $I_\alpha(x)$ are the Bessel functions of the first kind. The integral in~\eqref{id-2} 
can be efficiently evaluated numerically.


\end{document}